\documentclass{aastex}
\usepackage{emulateapj5}
\usepackage{epsf}

\newcommand{\HI}{H{\sc i}\ }
\newcommand{\MLstar}{\ensuremath{\Upsilon_{\star}}}

\renewcommand{\arraystretch}{.6} 

\shortauthors{de Blok, McGaugh \& Rubin}
\shorttitle{LSB rotation curves}

\begin{document}

\title{High-resolution rotation curves of LSB galaxies: Mass Models}

\author{W.J.G.~de~Blok\altaffilmark{1}}
\affil{Australia Telescope National Facility}
\affil{PO Box 76, Epping NSW 1710, Australia}
\email{edeblok@atnf.csiro.au}
\altaffiltext{1}{Bolton Fellow}

\bigskip

\author{Stacy S. McGaugh} 
\affil{Department of Astronomy, University of Maryland} 
\affil{College Park, MD 20742-2421, USA}
\email{ssm@astro.umd.edu}
 
\and 

\author{Vera C. Rubin}
\affil{Department of Terrestrial Magnetism}
\affil{Carnegie Institution of Washington }
\affil{5241 Broad Branch Rd., N. W.}
\affil{Washington, D. C. 20015, USA}
\email{rubin@dtm.ciw.edu}

\begin{abstract}

We present mass models for a sample of 30 high-resolution rotation
curves of low surface brightness (LSB) galaxies. We fit both
pseudo-isothermal (core-dominated) and Cold Dark Matter (CDM)
(cusp-dominated) halos for a wide variety of assumptions about the
stellar mass-to-light ratio.  We find that the pseudo-isothermal model
provides superior fits.  CDM fits show systematic deviations from
the data, and often have a small statistical likelihood of being the
appropriate model. The distribution of concentration parameters is too
broad, and has too low a mean, to be explained by $\Lambda$CDM. This
failing becomes more severe as increasing allowance is made for
stellar mass: NFW fits require uncomfortably low mass-to-light
ratios. In contrast, the maximum disk procedure does often succeed in
predicting the inner shape of the rotation curves, but requires
uncomfortably large stellar mass-to- light ratios.  The data do admit
reasonable stellar population mass-to-light ratios if halos have cores
rather than cusps.
\end{abstract}


\section{Introduction}

\subsection{LSB Galaxies}
Over the last five years the rotation curves of Low Surface Brightness
(LSB) galaxies and the constraints they impose on cosmological
theories have received much attention in the literature. An LSB galaxy
is usually defined as a disk galaxy with an extrapolated central disk
surface brightness $\gtrsim 1$ mag arcsec$^{-2}$ fainter than the
typical value for ``normal'' High Surface Brightness (HSB) spiral
galaxies \citep{freeman70}.  Colors, metallicities, gas-fractions, and
extensive population synthesis modelling all support the idea that LSB
galaxies are unevolved galaxies with low (current and past) star
formation rates
\citep[e.g.][]{vdh93,mcg_opt94,mcg_abund94,mcg_gas97,edb_phot95,vdhoek00,
bell_lsb00}. See \citet{lsbreview} for a review.

The observation that LSB and HSB galaxies follow the same Tully-Fisher
(TF) relation requires (in the conventional picture) that LSB galaxies
are dominated by dark matter (DM)
\citep{zwaan_tf95,spray_tf95,marc_phd97,edb_hsblsb96}. For reasonable
stellar mass-to-light ratios \MLstar\ low surface brightness implies
low stellar density. Yet, the extended, low surface density stellar
disks cannot be the major contributors to the dynamics in LSB
galaxies, as no shift in the zero-point of the TF relation with
surface brightness is observed.  This contrasts with the dominance of
the stellar population in HSB galaxies of similar luminosity.

The modest \MLstar\ values as implied by the blue colors and the
(baryonic) TF relation, together with the diffuseness of the stellar
disks make analyses of the DM distribution in LSB galaxies
less ambiguous than in HSB galaxies where the stellar component can be
significant even for fairly low \MLstar\ values.  LSB galaxies are
therefore ideal laboratories for measuring the distribution of dark
matter for comparison with predictions of theories of galaxy
formation.

For example, one of the results of numerical Cold Dark Matter (CDM)
simulations is a so-called ``universal halo mass density profile''
\citep{NFW96}, commonly known as a ``NFW profile.''  NFW (and all CDM)
mass density profiles are characterized by steep central cusps. This
is in contrast with the other commonly used ``classic''
pseudo-isothermal sphere halo model which is characterized by a
constant density core.  The parameters of the NFW mass density
distribution are related to the mass of the halo and the density of
the universe at the time of collapse and are therefore set by the
cosmology.  As these parameters can be determined from observations,
this opens the possibility of testing the NFW CDM model as well its
underlying assumptions.

A first analysis of LSB galaxy \HI rotation curves by \citet{edb_rot}
indicated that they did not rise as steeply as their HSB counterparts
of similar luminosity, contrary to CDM predictions.  The mass
distribution in LSB galaxies is more extended and of lower density
than in HSB galaxies \citep{edb_hsblsb96}.

Other results also indicate that the steep rotation curves implied by
CDM are hard to reconcile with the observed shallow rotation curves of
dwarf galaxies
\citep{moore94,floresprimack94,blais00,cote00}.  To explain this discrepancy
the possibility of redistribution of the (cuspy) DM due to
violent star-formation (thus creating the observed cores) was
sometimes raised, but this has been shown to be inconsistent with
other observational data \citep{maclow}.

\citet{mcg_nodm98} argued that the shapes of rotation curves of LSB 
galaxies were inconsistent with those predicted by the NFW
prescription.  This could not be explained by the effects of star
formation as the larger masses of LSB galaxies would require large
bursts in order to redistribute matter on large scales. Their
quiescent evolutionary history argues strongly against this
\citep{vdhoek00}.

This comparison with the CDM model is often dismissed because of the
limited resolution of the observed \HI curves.  The early \HI LSB
rotation curves were obtained using the {\sc vla} and {\sc wsrt} radio
synthesis telescopes. The relatively large beams of these instruments
resulted in rotation curves with only a limited resolution. 
\citet{edb_rot} did however show that for the best resolved cases
the effects of beam smearing were not strong enough to explain the
observed shallow curve as simply the result of a steep NFW model curve
affected by beam-smearing. Similar results were found for more
fashionable cosmologies, such as $\Lambda$CDM ($\Omega_m \sim 0.3,\
\Omega_{\Lambda}\sim 0.7$), though with smaller discrepancies.  

Even so, the theoretical debate now seems to have settled on halos
with cusps even steeper than NFW halos \citep{moore99}, thus worsening
the possible conflict between the data and the simulations.  From the
observational point of view the easiest and least ambigous way to test
the reality of these discrepancies is to measure high-resolution
rotation curves.

\subsection{Optical rotation curves}
Optical H$\alpha$ rotation curves of five LSB galaxies from the sample
of \citet{edb_bmh96} (BMH) were presented in \citet{SMT} (SMT). Though
SMT found that for two of the five galaxies the inner slopes of the
rotation curves were steeper than derived from the \HI
observations, this difference does not affect the BMH conclusion that
LSB rotation curves have shallower slopes than HSB rotation curves of
similar amplitude.  Because of these steeper slopes, SMT derive higher
maximum-disk \MLstar\ values (in some cases $> 10$), strengthening one
of the conclusions from \citet{edb_rot} that the maximum \MLstar\
values in LSB galaxies are too large to be accommodated by reasonable
star formation histories and Initial Mass Functions.  Such high values
are inconsistent with the existence of a baryonic TF-relation
\citep{mcg_barytf}.

A different approach was taken by \citet{frankvdb}. They attempted to
apply a rigorous correction for beamsmearing to the BMH
\HI data, and thus to derive the true ``infinite
resolution'' rotation curve.  They conclude that the data are not of
high enough resolution to accept or reject the NFW hypothesis with any
significance.  However, as they use a modified NFW profile with the
inner slope of the mass-density distribution as an (additional) free
parameter, it is not clear how significant this conclusion is.  The
usual 3-parameter rotation curve fits are already under-constrained;
adding another parameter does not improve the significance of the
results. Furthermore, in some cases they find such low values for the
inner slope that their NFW-halos effectively become
core-dominated. These halos do of course fit the data, but do not
occur in CDM simulations.

The general picture as derived from early observations of rotation
curves of LSB galaxies therefore still holds: LSB galaxies are
unevolved, low density galaxies, dominated by DM.  Their
rotation curves have shallower slopes than those of HSB galaxies of
similar amplitude, and the shape of the best-resolved LSB curves are
not necessarily consistent with the NFW rotation curve shapes.

\subsection{New data}

In this paper we present an analysis of high-resolution high-quality
hybrid H$\alpha$/\HI rotation curves of a sample of 30 LSB galaxies.
Of this sample 26 curves were taken from the large sample of 50 LSB
galaxies presented in McGaugh, Rubin \& de Blok (2001) (herafter Paper
I). In that paper an extensive description is given of the data, the
sample and reduction method.  We also refer to that paper for a
comparison of the new H$\alpha$ data with the BMH \HI curves.  We also
re-analyse the data for an additional 5 curves taken from \citet{SMT}.
In this paper we derive mass models under various assumptions for
\MLstar\ and fit these models both with NFW halos and
pseudo-isothermal halos. A similar analysis for a different set of
rotation curves of dwarf and LSB galaxies is given in de Blok \& Bosma
(2001).

In Section 2 we discuss the sample, and discuss the derivation of the
rotation curves. We also show internal and external comparisons of the
data and discuss possible systematics.  In Section 3 we discuss the
various mass models. Section 4 contains the results of the model
fitting.  Section 5 discusses the implications for the various halo
models. In Section 6 we turn our attention to the maximum disk and a
summary is given in Section 7.

When using absolute distances we have used a Hubble constant $H_0 =
75$ km s$^{-1}$ Mpc$^{-1}$.

\section{The data}

\subsection{Sample and raw data}

The data and reduction methods have been extensively described in
Paper I. In summary, we use long-slit major axis spectra taken with
the 4-m telescope at Kitt Peak in June 1999 and February 2000 and the
100$''$ telescope at Las Campanas in November 1998.  Velocities were
derived from the intensity weighted centroid of the H$\alpha$ and
[N{\sc ii}] lines.

As the aim of this excercise is to derive mass models which can yield
significant constraints on the distribution of DM we select
only the 26 high quality galaxies from Paper I.  We split this
high-quality sample in two sub-samples.  Sample I contains LSB
galaxies from BMH and
\citet{vdh93} for which a full set of photometry and \HI data 
is available.  In Sample I we also include the five galaxies presented
in SMT. For these galaxies \HI and optical photometry are taken
from BMH.  Tables~\ref{fgalsample} and \ref{esosample} contain a full
list of the galaxies analysed here, along with some of their global
parameters.

Sample II consists of ESO-LV and UGC LSB galaxies, for which an
optical rotation curve is available but no optical or
\HI photometry.

For the galaxies in Sample I, \HI observations are available which
often extend to larger radii than the H$\alpha$ data.  To make the
best use of both types of data, we have constructed hybrid rotation
curves.  These consist of the H$\alpha$ data over the range of radii
where available, and 21-cm data to define the outermost points.  No
attempt has been made to ``average'' the different types of data:
H$\alpha$ is given precedence over the range of radii where it is
available.

\subsection{Derivation of the smooth curves}

One of the main assumptions made when deriving mass models from
rotation curves is that the gas and stars trace circular orbits in an
axisymmetric potential.  Though the shape of the optical rotation
curves in Paper I is well-defined, the scatter between individual
datapoints means we cannot simply use the raw rotation curves to
estimate the radial run of the gravitational potential.
For this one needs a smooth
curve which retains real small-scale details, but without the
observational scatter.

The method most often used to produce these smooth curves is to fit
splines to the data. Here we use have used a robust version of this
procedure (local regression, see Loader 1999).  The smooth curves were
rebinned to a bin width of 2$''$.  The error bars in the rebinned data
points consist of two components: one due to observational errors
caused by the measurement uncertainties in the individual raw
datapoints (for this we use the average weighted measurement error in
each bin), and an additional component caused by differences between
approaching and receding sides and non-circular motions (which we
define as the difference between the weighted mean raw velocity and
the velocity implied by the spline fit at that radius). For the final
error estimate these two uncertainties were added quadratically.

For some high S/N data points the error bars become unrealistically
small (sometimes less than 1 km s$^{-1}$). This has no physical
significance and simply tells us that the profile centroids
were well-determined.  These small error bars can however easily
dominate any model fit and can severely bias $\chi^2$ values or
goodness-of-fit parameters.  For this reason, and as the observational
and physical uncertainties (slit position, streaming motions) make it
difficult to determine a physically meaningful rotation velocity with
an accuracy of more than a few km s$^{-1}$, we have imposed a minimum
error on each point of 4 km s$^{-1}$ (before inclination
correction). The curves were corrected for inclination using the
values given in Tables~\ref{fgalsample} and \ref{esosample}.

\begin{figure} 
\begin{center} 
\epsfxsize=0.95\hsize
\epsfbox{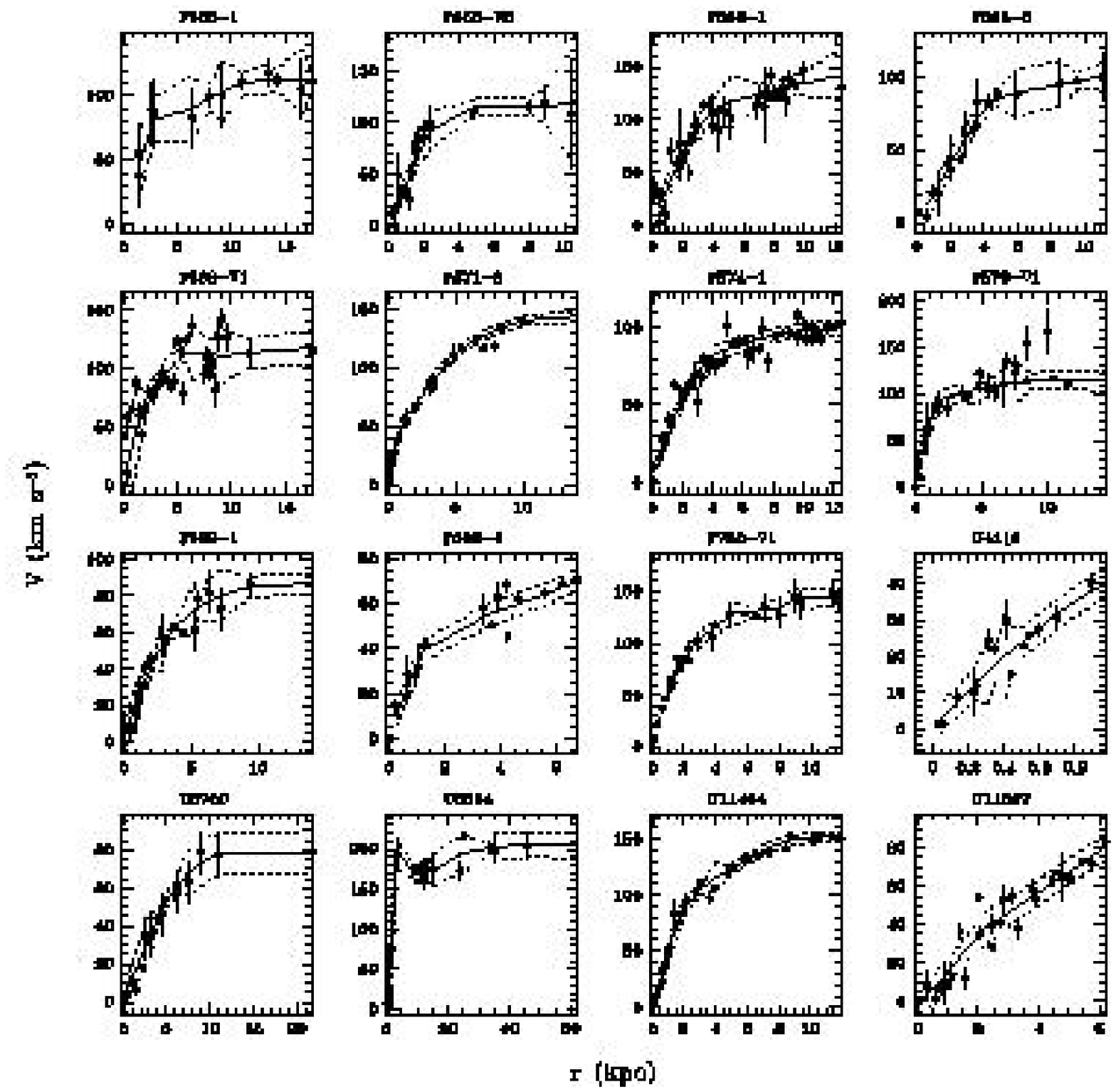}
\figcaption[fig1a.ps]{Comparison of the raw hybrid rotation curves (black
dots) with the smooth curves (full lines). The derived uncertainties
in the smooth curves are indicated by the dotted lines. 
\label{smoraw}}
\end{center}
\end{figure}

The end result is a smooth representation of the data, which is
reproducible and as objective as possible, to use as input for the
mass models. Figure~\ref{smoraw} show overlays of both the raw hybrid
curves and the smooth versions. It is easy to verify that no
systematic differences in slope or shape have been introduced.  The
error bars in the smooth curves are also a good representation of the
uncertainties in the underlying raw data.

Table~\ref{electable} contains the hybrid smooth rotation curves.  For
each galaxy we list the radii in arcsec, as well as in kpc, together
with the observed rotation velocities and the uncertainties in these
values. Also included are the rotation curves for the gas component
(already included is the factor 1.4 mass scaling for He), the disk
component (values listed assume $\MLstar(R) = 1.0$), and where
applicable the bulge component (also for $\MLstar(R) = 1.0$.

\subsection{Comparing the smooth curves}

\begin{figure} 
\begin{center} 
\epsfxsize=0.8\hsize
\epsfbox{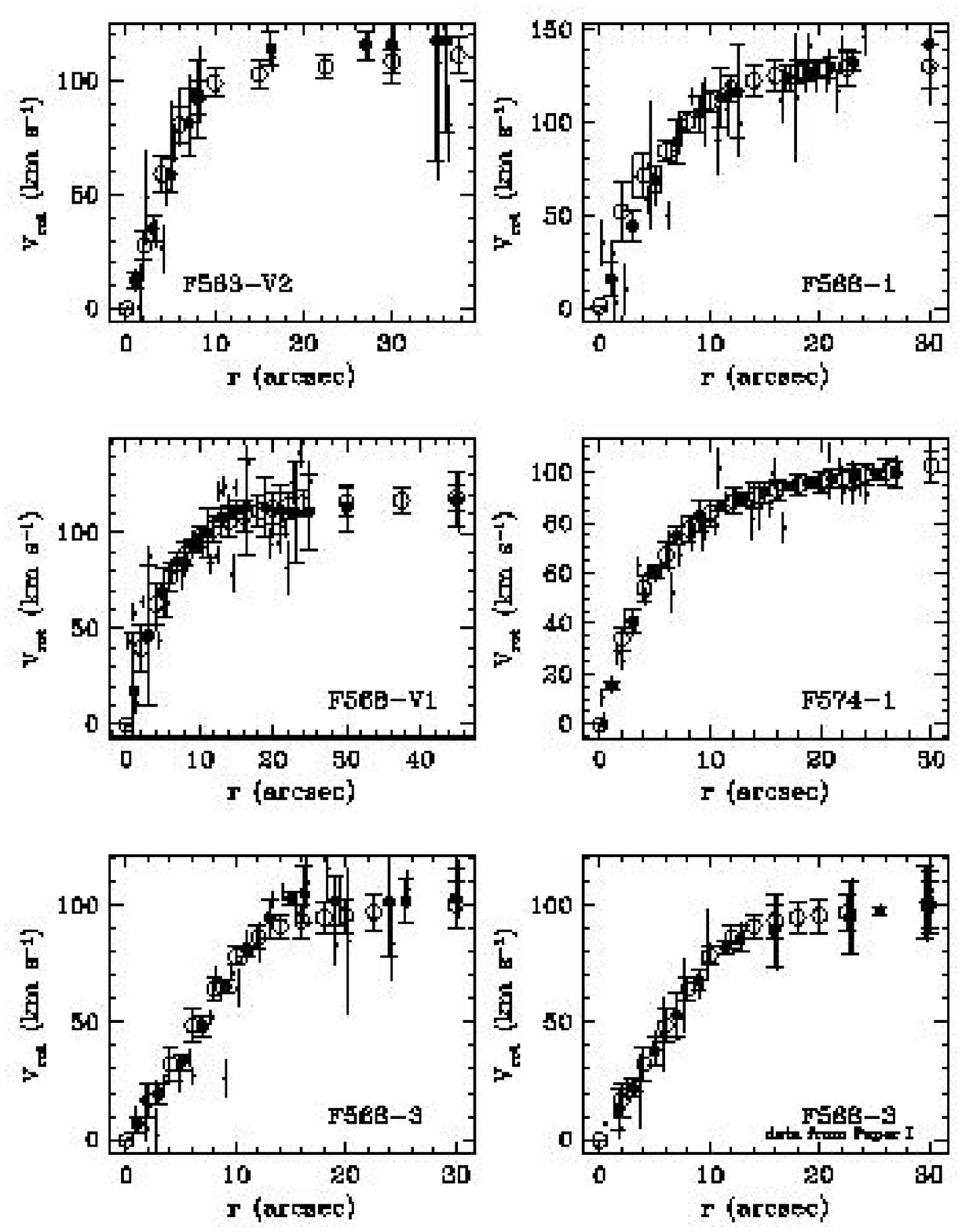} 
\figcaption[fig2.ps]{Comparison of our
analysis of the STM data with their resampled rotation curves.  The
top four panels and the bottom left panel show the raw data from STM
(grey small dots), their resampled and smoothed rotation curves (open
circles), and our local regression fits to the same data (filled large
circles). The raw data have been offset by +0.2$''$ to avoid overlap
with the binned data. The bottom right panel shows the raw data for
F568-3 taken from Paper I, along with the STM model (identical to the
model shown in the bottom-left panel) and our resampled rotation curve
based on the data from Paper I. Small differences between the various curves
are discussed in the text.\label{swcomp}}
\end{center}
\end{figure}

As noted above, we have included the five LSB galaxies presented in
SMT in our Sample I. As SMT show their raw data and derived smooth
curves, we can compare both sets of smooth rotation curves to
investigate possible systematics in our respective methods.  This is
done in Fig.~\ref{swcomp}. It is clear that the correspondence between
both velocities and errorbars is good and the differences are
minor. In most cases (F568-1, F568-V1 and F574-1) both sets agree at
(better than) the 1$\sigma$ level. Small remaining differences are
usually caused by a slightly different estimate of the velocities in
sparsely sampled parts (F563-V2 and F568-3).  The SMT curve for
F563-V2 (Fig.~\ref{swcomp}, top-left) is slightly higher than our
curve in the inner parts, but falls below ours in the outer parts.
The sparseness of optical data points in the outer parts and different
interpretations of the continuity between \HI and optical data are
probably the main cause of this difference.  For F568-3
(Fig.~\ref{swcomp}, bottom) we find a small ($\lesssim 1\sigma$)
systematic difference between both smooth curves. This galaxy has been
measured independently by SMT and by us (Paper I).  These raw data
sets agree in detail, and the difference must therefore be due to a
slightly different interpretation of the sparse raw data. In summary,
the smooth curves we present here give a good and reproducable
representation of the data.

\section{Mass models}

In order to find the signature of the dark halo one needs to model
the observed rotation curve using a number of separate dynamical
components, described below. 

\subsection{Stellar component} To model the stellar disk, the $R$-band
photometry presented in \citet{edb_phot95} was used.  The rotation
curve of the disk was computed following \citet{cas93} and
\citet{beeg_phd}. The disk was assumed to have a vertical sech$^2$
distribution with a scale height $z_0 = h/6$ \citep{kruit81}.  The
rotation curves of the stellar component were resampled at the same
radii as the smooth curves. We assume \MLstar\ is constant with
radius. While one expects some modest variation in $\MLstar$ with
radius \citep{dejong_grad}, the color gradients in LSB galaxies tend
to be small, so this effect is not likely to be significant.

\subsection{Gas disk} The \HI surface density profiles presented in BMH
and \citet{vdh93} were used. They were scaled by a factor of 1.4 to
take the contribution of helium and metals into account. Their
rotation curve was derived assuming the gas was distributed in a thin
disk. The gas rotation curves were resampled at the radii of the
smooth observed rotation curve.

\subsection{Dark halo} 

The dark halo component differs from the previous two in that we are
interested in parametrizing this component assuming some fiducial
model. The choice of this model is the crux of most of the DM
analyses in the literature, and many models exist.  These can be
broadly distinguished in two groups: halo models with a core, and halo
models with a cusp. An example of the first category is the
pseudo-isothermal halo, an example of the latter the CDM NFW halo.

As one of the goals of this paper is to assess the relevance of either
category to the high-resolution LSB galaxy rotation curves, we will
present models derived using both models. We do realize there are many
intermediate models described in the literature that probably can fit
our data equally well. However, our goal here is simply to see where
the data lead us: is there a preference for models with a core or
with a cusp? We now describe the details of both models.

\subsubsection{Pseudo-isothermal halo}

The spherical pseudo-isothermal halo has a density profile
\begin{equation} 
\rho_{ISO}(R) = \rho_0 \Bigl[ 1 + \Bigl(
{{R}\over{R_C}} \Bigr)^2 \Bigr]^{-1}, 
\end{equation} 
where $\rho_0$ is
the central density of the halo, and $R_C$ the core radius of the
halo.  The corresponding rotation curve is given by
\begin{equation} V(R) = \sqrt{ 4\pi G\rho_0 R_C^2 \Bigl[ 1 -
{{R_C}\over{R}}\arctan \Bigl( {{R}\over{R_C}} \Bigr) \Bigr] }.
\end{equation}
The asymptotic velocity of the halo, $V_{\infty}$, is given by
\begin{equation} 
  V_{\infty} = \sqrt{ 4 \pi G \rho_0 R_C^2 }.
\end{equation}
To characterize this halo only two of the three parameters
$(\rho_0, R_C, V_{\infty})$ are needed, as equation (3) determines the
value of the third parameter.  

\subsubsection{NFW halo}

The NFW mass density distribution takes the form
\begin{equation}
\rho_{NFW}(R) = \frac{\rho_i}{\left(R/R_s\right)
\left(1+ R/R_s\right)^2}
\end{equation}
where $R_s$ is the characteristic radius of the halo and
$\rho_i$ is related to the density of the universe at the time of collapse.  
This mass distribution gives rise to a halo rotation curve
\begin{equation}
V(R) = V_{200} \left[\frac{\ln(1+cx)-cx/(1+cx)}
{x[\ln(1+c)-c/(1+c)]}\right]^{1/2},
\end{equation}
where $x = R/R_{200}$.
It is characterized by a concentration parameter $c =
R_{200}/R_s$ and a radius $R_{200}$. These are directly related to
$R_s$ and $\rho_i$, but are used instead as they are a convenient way
to parametrize the rotation curve.  The radius $R_{200}$ is the radius
where the density contrast exceeds 200, roughly the virial radius
\citep{NFW96}. The characteristic velocity $V_{200}$ of the
halo is defined in the same way as $R_{200}$.  These parameters are
not independent and are set by the cosmology.

\subsection{Mass-to-light ratios and weighting}

One of largest uncertainties in any mass model is the value of 
\MLstar. Though broad trends in \MLstar\  have been measured and modelled 
\citep[e.g.][]{botje97,bell_pops}, the precise value for an individual galaxy
is not well known, and depends on extinction, star formation history,
Initial Mass Function, etc.  Rotation curve fitting is a problem with
too many free parameters
\citep{maxdisk86,lake89} and some assumptions regarding \MLstar\ must
be made.  We therefore present disk-halo decompositions using four
different assumptions for $\MLstar$ for the galaxies in Sample I. For
the galaxies in Sample II only the minimum-disk model is presented.

{\bf Minimum disk.} This model assumes that the observed rotation
curve is due entirely to DM. This gives an upper limit on how
concentrated the dark mass component can actually be and is the
version of minimum disk preferred in the CDM literature.

{\bf Minimum disk + gas.}  The contribution of the atomic gas (H{\sc
i} and He) is taken into account, but \MLstar\ is assumed to be
zero. This is the classical definition of minimum disk as used in the
\HI rotation curve literature.

{\bf Constant \MLstar.} Here \MLstar\ is set equal to a constant value
based on an Initial Mass Function and a star formation history
appropriate for LSB galaxies.  For the range in color $0.4 < B-V <
0.65$ which LSB galaxies normally exhibit \citep{edb_phot95} a value
$\MLstar(R) = 1.4$ is a good estimate.  For example, using the
\citet{ch_bruz} model with constant star formation rate and Salpeter
IMF, we find that $\MLstar(R)=1.4$ corresponds to $B-V=0.46$. The {\sc
pegase2} model (Rocca-Volmerange, priv.\ comm.)  gives a value
$B-V=0.38$, whereas the model by \citet{cole00} yields $B-V=0.67$. The
models by \cite{bell_tfproc} give values around $B-V \simeq 0.6$.  The
value $\MLstar=1.4$ is thus actually at the ``light-weight'' end of
the plausible range, but this was deliberately chosen in order to give
maximum opportunity for the cuspy NFW models to fit the data.  We
realize that the values derived here should not be regarded as
definitive: changes in the IMF model used or different estimates for
internal extinction can lead to different values. However, here we 
attempt to derive 
a conservative estimate for \MLstar\ based
on the observed properties of the stellar population.  Further
(upward) refinement of the $\MLstar$ value is thus more likely to
cause more problems for NFW fits.

{\bf Maximum disk.} The rotation curve of the stellar component is
scaled to the maximum value allowed by the (smooth) rotation curve,
but with the restriction that the DM density is required to be
positive at all radii (thus avoiding a so-called ``hollow halo'')
\citep{maxdisk86}.  Because of the different DM distributions
that we test (core and cusp) this can ocassionally lead to maximum
disk values that differ slightly for each of the two models. A more
extensive description is given in \S 6.

Each of the rotation curves was fitted using the {\sc gipsy} task {\sc
rotmas}. The program determines the best-fitting combination of $R_C$
and $V_{\infty}$ (for the pseudo-isothermal halo) or $c$ and $V_{200}$ (for
the NFW halo), using a least squares fitting routine.  We assigned
weights to the data points inversely proportional to the square of
their uncertainty.  Additional checks were made with other fitting
programs to check the results of the fits (discussed in Sect.~4.1
below). We also refitted the smooth rotation curves presented in SMT
and were able to reproduce the numbers given in their Table~2.

\section{Results}

Tables \ref{nfwtable} and \ref{nfwesotable} give the results of the
model fitting using the NFW halo model. Tables \ref{isotable} and
\ref{isoesotable} show the results for the pseudo-isothermal halo.
Figure~\ref{nfwiso} presents the results of the NFW halo and
pseudo-isothermal halo mass modelling for each galaxy side-by-side.  The two
leftmost columns show the results for NFW halo fitting.  The two
rightmost columns show the pseudo-isothermal halo fitting results.

The first and third columns show the best-fitting models. The rotation
curves of the gas are shown as dotted lines, those of the stellar disk
component as short-dashed lines. In two cases (U6614 and F571-8) a
significant bulge was present which was modelled separately; this is
shown as the dot-dashed line (see also Section 4.3). The resulting
halo rotation curve is shown as the long-dashed line. The final total
model curve is drawn as the full line.  In each of the model panels we
also give the reduced $\chi^2$ of the fit and the chance $p$ that the
data and the model could result from the same parent
distribution. This probability was derived using a simple $\chi^2$
test; it is an indicator for the compatibility of the data and the
model chosen to describe it. Values $p>0.95$ indicate that the data
and the model are a good match. Values $p<0.05$ indicate that the
model is incompatible with the data, and that better models can be
found.

We show two best-fitting values. One as found by the linear fit of the
{\sc gipsy rotmas} task (indicated as a ``plus''-cross (+)), and one
found by finding the minimum in the plotted logarithmic parameter
space (indicated by a ``times''-cross $(\times)$).  These two are
identical, except when extreme parameter values occur (usually during
maximum disk fits) and numerical precision of the fitting routine
starts to play a role. This effect is visible in the bottom-right
corners of the NFW contour plots, where the very large values of
$V_{200}$ in combination with the small values of $c$ cause
increasingly ragged contours. A large difference between these two
best values therefore indicates that the fit should not be regarded as
definitive. Indeed, in a number of these cases (indicated in the
Tables as italic numbers) the fitting routine was unable to determine
a valid solution, and an indicative value had to be chosen by
hand. This happened mostly with the NFW models. One of the reasons for
this is that the inner parts of the rotation curves can be well
described by $V(R) \sim R$, whereas the NFW model has the form $V \sim
R^{1/2}$. To accommodate the model, the fit tries to stretch out the
NFW curve (resulting in small $c$ and high $V_{200}$) in order to make
it
\emph{look} linear.

The second and fourth columns of panels in Fig.~\ref{nfwiso} show the
1$\sigma$ (thick contour) and $(2,3,4,5)\sigma$ (thin contours)
probability contours of the halo parameters in logarithmic space.  The
reason for choosing a logarithmic representation is that the $\chi^2$
distributions for the NFW halo parameters often show extended tails
towards very small $c$ and/or large $V_{200}$.  For comparison,
Fig.~\ref{linchi2} shows a representative example of the
$\sigma$-contours for the minimum disk model of F583-4 plotted in
linear $c$-$V_{200}$ space.

It is important to realize that  the $\sigma$ contours are plotted 
\emph{with respect to the minimum $\chi^2$}. That is, 
existence of a narrow distribution only means that the minimum
$\chi^2$ is well defined. It does \emph{not} imply that the fit is
good in an absolute sense. For that one needs to refer to the value of
the reduced $\chi^2$ itself or the probability $p$ that the data and
model are compatible. There are many cases where the NFW model is not
a good fit, making it difficult to plot absolute likelihood contours.

Finally, in the NFW contour plots in the second column we show the
range of $c$ and $V_{200}$ values for the currently popular
$\Lambda$CDM cosmology as derived from numerical models
\citep{NFW97} (see \S 5.3). 
The cross-hatched and single hatched grey areas shows the expected
1$\sigma$ and 2$\sigma$ logarithmic scatter in $c$ (where $\sigma_{c}
= 0.18$) as found in numerical models by
\citet{bul99}. Independent simulations by \citet{jing99} find a much
smaller logarithmic scatter of $\sigma_{c} = 0.08$.  The latter do of
course put much stronger constraints on the NFW results.  For the sake
of clarity, however, and to give the NFW model as much chance as
possible we adopt the larger estimate of the scatter in $c$ of
\citet{bul99}.

For the pseudo-isothermal halo we show contours of constant central density
$\rho_0$.  The contours represent from top to bottom $\rho_0 = $ 0.1
(dotted), 1 (dashed), 10, 100, 1000 (dotted) $\times 10^{-3}\
M_{\odot}$ pc$^{-3}$.

\subsection{Weighted versus uniform}

\begin{figure} 
\begin{center} 
\epsfxsize=0.8\hsize \epsfbox{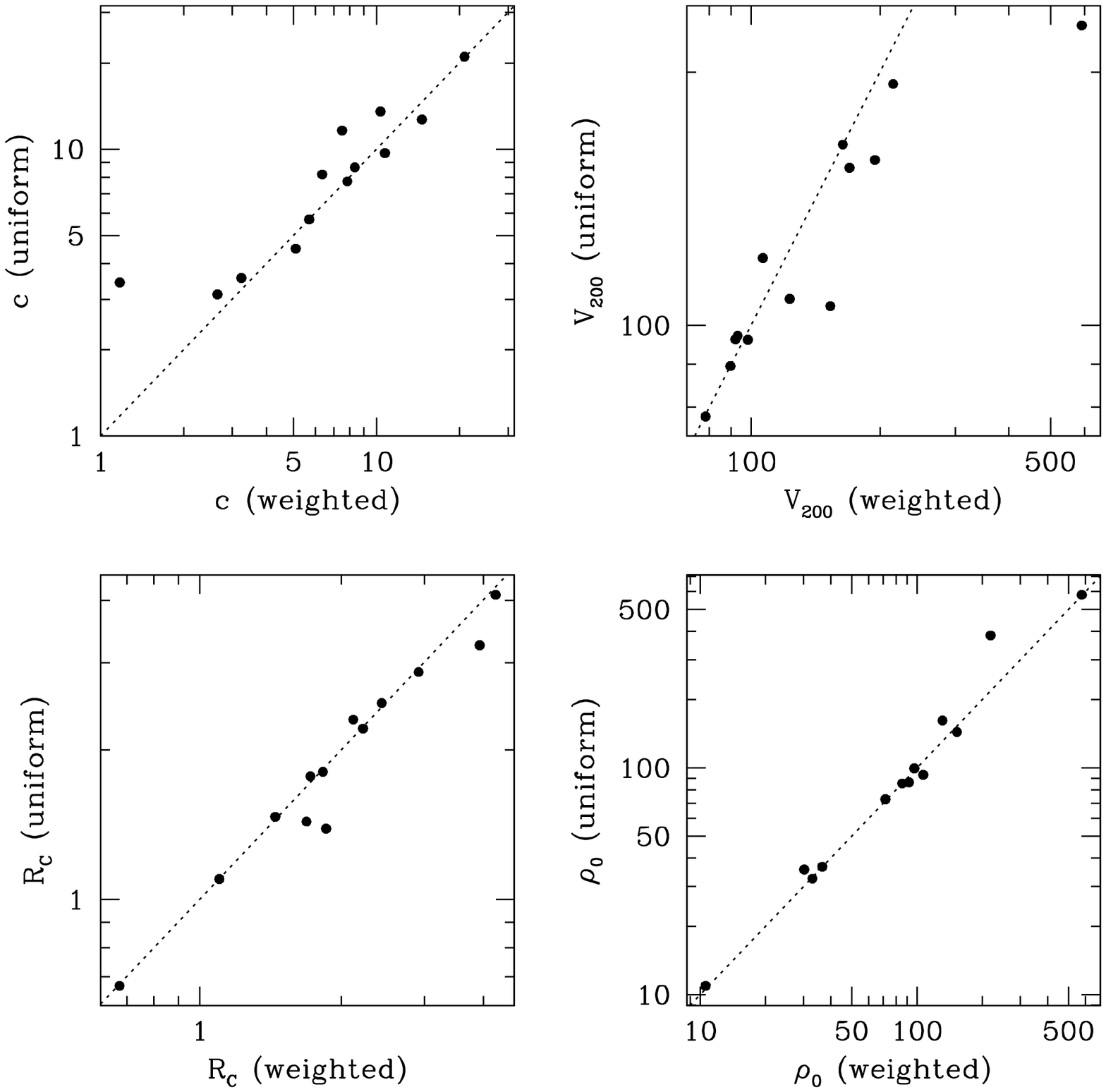}

\figcaption[fig5.ps]{Comparison of the NFW halo parameters $c$ and $V_{200}$ 
and the pseudo-isothermal halo parameters $R_C$ and $\rho_0$, as derived
using a weighted model fit, using the inverse variance as weight, and
uniform errorbars. There is good agreement. The somewhat increased
scatter at extreme values is not significant, as the shape of the
model in that area of parameter space is fairly insensitive to the
precise values.
\label{compweightunif}}
\end{center}
\end{figure}

To investigate how stable the derived halo parameters are with respect
to the precise definition of the errorbars, we have re-derived the
models assigning uniform and equal weights to all data points.  Though
we do not list the latter values here, we show in
Fig.~\ref{compweightunif} a comparison between the two sets of
parameters.  It is clear that these agree well, showing that the
results presented here are robust against the precise definition of
the errorbars.

\subsection{Mass models: smooth and raw}

Has the procedure used to derive the smooth rotation curves affected
some of the model results? We established in Sect.~2.4 that this
procedure introduced no systematic differences between the smooth
curves and the raw data.  Here we test this again by checking whether
the smooth curves give the same fit results as the raw data.

\begin{figure} 
\begin{center} 
\epsfxsize=0.8\hsize \epsfbox{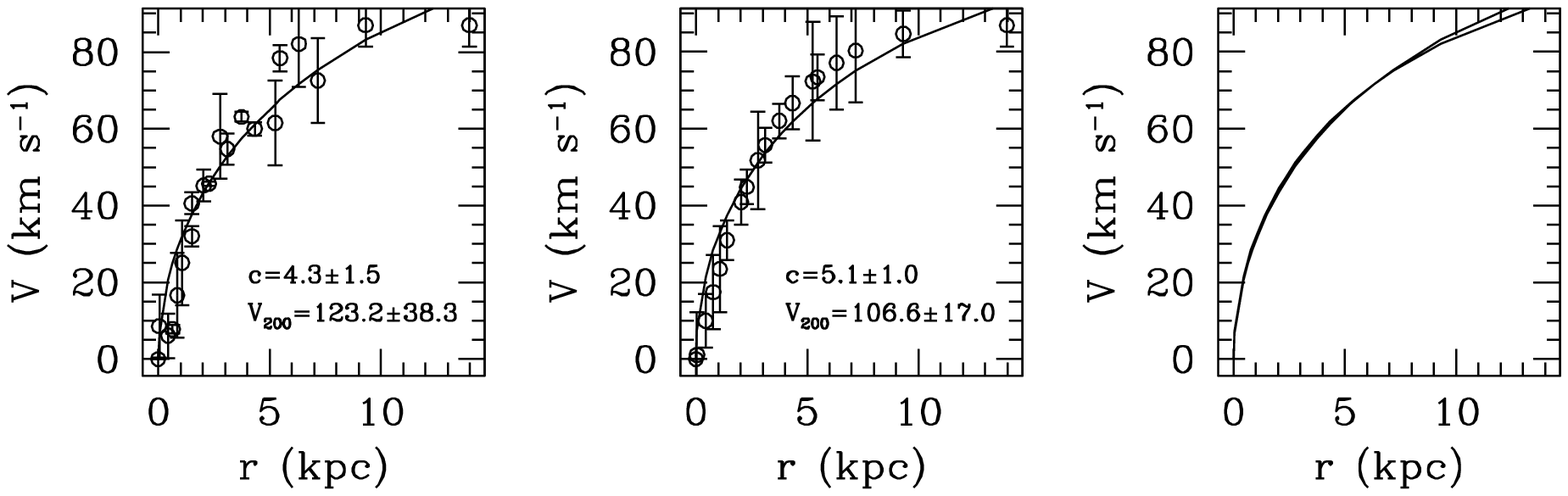}
\figcaption[fig6.ps]{Comparison of NFW minimum disk fits to the raw rotation
curve of F583-1 (left panel) and the smooth curve (middle panel). The right
panel compares the two fits, which are virtually identical. The fit parameters
shown in the panels agree within their errors.\label{f5831comp}}
\end{center}
\end{figure}

As the NFW model is more sensitive to changes in the inner slope than
the pseudo-isothermal model we will use the former in our checks.  We first
fit NFW minimum disk models to the smooth and the raw curve of F583-1,
as a representation of the data from Paper I. Both fits are presented
in Fig.~\ref{f5831comp}, where we have imposed a minimum error of 4 km
s$^{-1}$ on the raw data to make the errorbars consistent with the
smooth curve. It is clear that the two fits are identical within the
errorbars. Similar results are obtained using other curves from Paper
I.

As a second test we evaluate the SMT data. We have fitted several
minimum disk NFW models to each of the SMT galaxies.  We have fitted
the SMT raw data, the smooth curve presented in SMT, and our smooth
curve derived from the raw SMT data.  These fits were done
independently by two of us using independent fit codes on the smoothed
(dB) and unsmoothed (McG) data.  Table~\ref{comptable} lists the
derived parameter values. For comparison, we also list the results for
our own independent observation of F568-3 from Paper I.  In
Fig.~\ref{nfwswaters} we compare the $c$ values derived for each
galaxy.

We see that the galaxies for which our smooth curves and the ones
presented in SMT agree, also have similar model parameters, which
agree with those derived from the raw data (F574-1 and F568-V1).  In
the other three cases (F563-V2, F568-1, F568-3) the $c$-values derived
from our version of the smooth curves agree with those derived from
the raw data, whereas the SMT $c$-values are higher.  It is important
to keep in mind that even though the formal fit values show a large
discrepancy, the rotation curves themselves only show very subtle
differences (Fig.~\ref{swcomp}). This illustrates the importance of
having high-accuracy rotation curves of a large sample.  In the
following we only consider our smooth versions of the SMT data.

In summary, we believe that the results from our smooth curves are not
systematically different from the raw data. As stated before, we
prefer to use the smooth curves as these are more evenly sampled, and
prevent the occurence of imaginary halo masses which can arise when
the occasional (raw) data point happens to scatter below the rotation
velocity of the disk alone.

\subsection{Remarks on individual galaxies}

{\bf F563-1} For this galaxy independent observations are available
from de Blok \& Bosma (2001). See Paper I for a comparison. Note that
the observed curve differs significantly from the ``beam-smearing
corrected'' model presented in \citet{frankvdb}. The model presented
there shows an almost flat rotation curve over most of the radial
range, which clearly disagrees with the new data. Beam-smearing
corrections are not infallible.

{\bf F563-V2} This is our version of one of the SMT curves. This curve
does significantly worse at fitting NFW than a pseudo-isothermal halo.  The
systematics seen here are typical for many of the NFW fits: the inner
parts are overestimated; the model then underestimates the middle
parts, and shoots up again in the outer parts. For this galaxy no
$R$-band photometry is available and we have used $B$ band photometry
from \citet{mcg_opt94}.  Assuming $B-R=0.9$, which is the typical
color for an LSB, this yields a value for the constant $\MLstar$ case
of $\MLstar(B)=1.1$. The maximum disk NFW model fits significantly
better than the other NFW models. It is however not compatible with
cosmological predictions from the numerical models.

{\bf F568-1} This is another of the curves presented in SMT. 
The systematics of
overestimating the inner part, underestimating the middle, and
overestimating the outer velocities again are also present here. Again
maximum disk is the best of the NFW models, which is another way of
saying that the shapes of the (inner) rotation curves are more like
that expected for the stars (albeit with the wrong $\MLstar$).

{\bf F568-3} This is a well determined curve, for which there are
several consistent independent measurements (see Paper I).

{\bf F571-8} This is the only edge-on galaxy in Sample I, so we are
concerned about optical depth and projection effects in the optical
data. (The \HI data are not used for this galaxy.) These effects could
cause us to measure the rotation velocity at a ring where the optical
depth becomes unity. Recently,
\citet{matthews_edgeon} have used radiative transfer models to
investigate optical depth effects on rotation curves in edge-on LSB
galaxies, and they conclude that these effect are likely to be small
due to the low dust content in LSB galaxies. \citet{bosma92}, in a
comparison of the optical and \HI curves of the edge-on galaxy N100,
also finds that late-type galaxies tend to be transparant, even when
seen edge-on.  Nevertheless, we cannot exclude the possibility that
the shape of the optical curve is affected. If this is so then
\emph{in this case} the mass model will change.
For the constant \MLstar\ and maximum disk cases we have added an
exponential bulge with $\MLstar{}(R)_{\rm bulge} = 0.5$. See BMH for a
description of the bulge-disk decomposition.

{\bf F574-1} Another SMT curve. While the pseudo-isothermal halo fits for
this galaxy are good, the NFW fits show the by now familiar discrepancy: too
steep in the inner part, underestimating the middle, and rising too
quickly in the outer parts. Maximum disk NFW provides a good fit,
albeit with low $c$ and high \MLstar.  F574-1 was the worst case of
beam-smearing from the \HI sample, but the increase in the initital
rate of rise of the rotation curve found optically does not really
help NFW. The optical data imply a `cusp' slope ($\rho \propto
r^{\alpha}$) of $\alpha=-0.49 \pm 0.26$ \citep{paper3}, still well
short of the NFW value $\alpha=-1$. This is the limit in the minimum
disk case; if allowance is made for stellar mass, a value even
closer to a constant density core is required.

{\bf F583-1} A well resolved and well-observed curve that shows the
NFW over/under/over-fit discrepancy.  For all assumptions about
stellar mass, $\chi^2_{\rm ISO} \ll
\chi^2_{\rm NFW}$. This galaxy strongly prefers a halo with a constant 
density core over one with a cusp, a conclusion which has not changed
from \citet{mcg_nodm98}. Only a substantial change in the shape of the
rotation curve would alter this conclusions, which would require a
large \emph{systematic\/} error. Beam smearing can no longer be
invoked as the cause of such a systematic error now that this object
has been resolved to sub-kpc scales.

{\bf U5750} This curve was observed both by us and de Blok \& Bosma
(2001), and the two data sets show good agreement (see Paper I).  This
curve is difficult to model with a standard NFW profile, but the
pseudo-isothermal model provides a good fit.  The outermost point is taken
from the \HI curve, and provides an important constraint for the NFW
model. Without this point the fit produces $c\sim 0$ and an impossibly
large $V_{200}$, a result of the fitting program trying to make $V(R)
\sim R^{1/2}$ look like $V\sim R$.

{\bf U6614} This is the only giant LSB in Sample I. The analysis is
complicated by the presence of a dominant bulge, which we have
modelled as an exponential spherical bulge with $h=3.0''$ and
$\mu_0(R)=18.4$ mag arcsec$^{-2}$. The disk has parameters $h=19''$
and $\mu_0(R)=21.3$ mag arcsec$^{-2}$. The rapid rise and subsequent
dip in the rotation velocity at small radii clearly suggest the
dominance of the bulge in this giant LSB galaxy. We have assumed the
bulge to be maximal at $\MLstar{}(R)=3.7$. As the bulge accounts for
most of the rotation velocity in the inner parts, this seems to argue
against cuspy halos in giant LSB galaxies.

\section{Discussion}

\subsection{NFW and pseudo-isothermal: a comparison}

\begin{figure} 
\begin{center} 
\epsfxsize=0.8\hsize \epsfbox{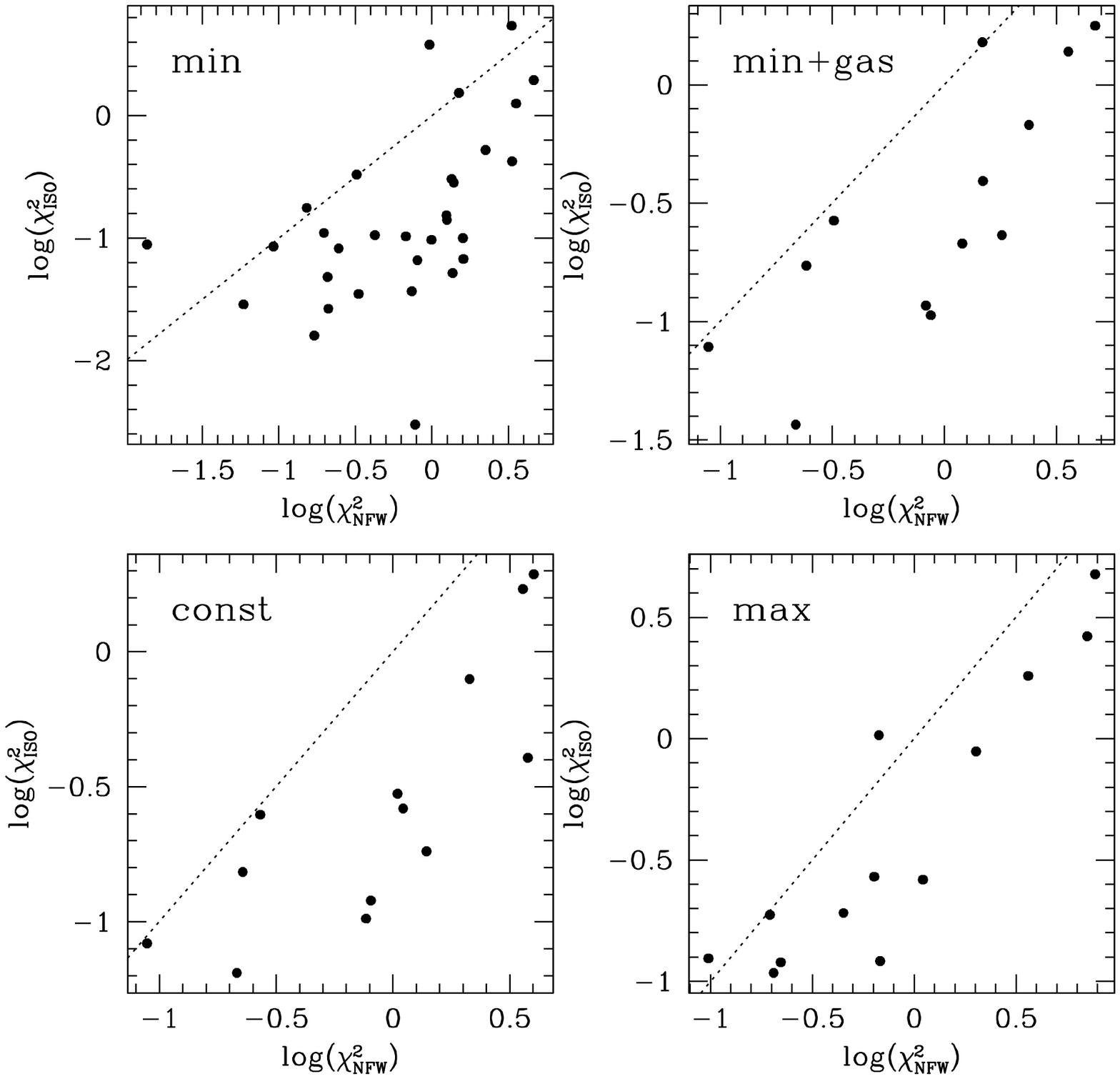}
\figcaption[fig8.ps]{Comparison of the reduced $\chi2$ values using NFW
and ISO halos, using the four assumptions for \MLstar\ as described in
the text. Note that the axes have logarithmic scales. The dotted lines
are lines of equality.\label{compchi2}}
\end{center}
\end{figure}

The pseudo-isothermal halos generally provide better fits than the NFW
halos. In Fig.~\ref{compchi2} we compare the reduced $\chi^2$ values
for the four different \MLstar\ cases.  For the minimum
disk case we plot Samples I and II, for the other cases only Sample I
is plotted. It is clear that the large majority of the curves
presented here is best fitted by a pseudo-isothermal halo. This holds true
even in the maximum disk case where one might naively expect the
dominance of the optical disk to wipe out any discrepancies of a
particular halo model (though perhaps not for LSB galaxies).

Another way of comparing the models is given in
Table~\ref{veratable}. This lists the number of galaxies in Sample I
that have good ($p>0.95$) or bad ($p<0.05$) fits for each of the two
models.  Here again it is clear that the pseudo-isothermal model
performs much better, for every assumption of \MLstar.  These
statements do not depend on the errors. If we double (or halve) the
size of the error bars, $\chi^2$ will change for both halo cases, but
will always remain less for the pseudo-isothermal case. To alter this result
would require systematic changes to the shapes of all the rotation
curves.

\begin{figure} 
\begin{center} 
\epsfxsize=0.8\hsize \epsfbox{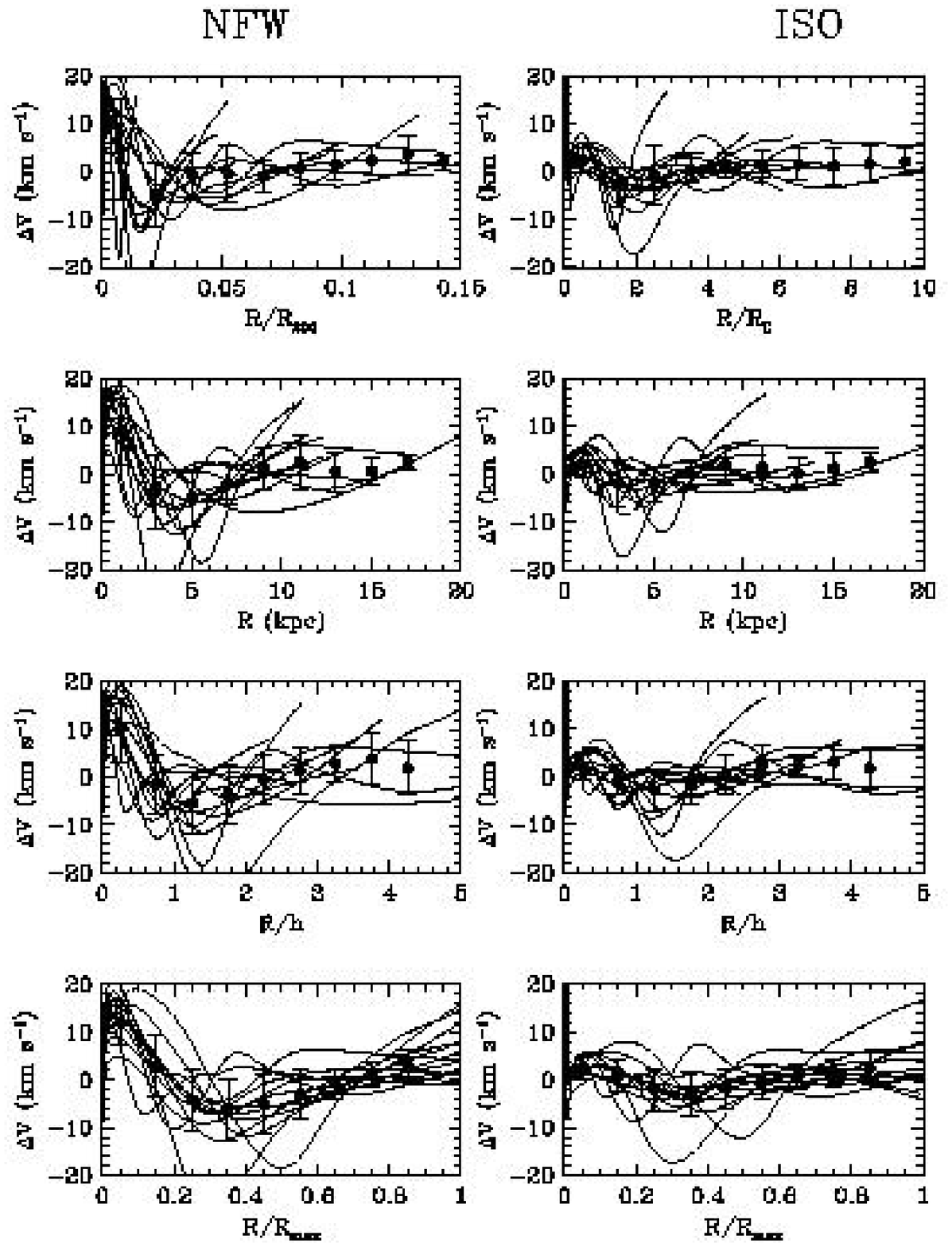}
\figcaption[fig9.ps]{Comparison of the residuals $\Delta V = 
V_{model}-V_{obs}$ for Sample I (minimum disk), plotted against number
of halo scale radii (first row), absolute radius (second row), number
of optical disk scale lengths (third row) and fraction of maximum
radius of rotation curve (fourth row).  The left panels show residuals
using NFW halo models, the right panels show the pseudo-isothermal halo
case. Also shown are the average residuals and standard deviations
(filled dots). The residuals at small radii are much larger for the
NFW model than for the ISO model. The low-level systematic residuals
that are also apparent for the pseudo-isothermal halo, probably tell us that
real halos are subtly different from pseudo-isothermal.
\label{residuals}}

\end{center}
\end{figure}

Fig.~\ref{residuals} shows the residuals of the best-fitting minimum
disk models versus the observed data. Residuals are plotted against
halo scale size ($R_{200}$ for NFW and $R_C$ for pseudo-isothermal halos),
radius in kpc, number of optical disk scale lengths and fraction of
maximum radius of the rotation curve. As described in the previous
section, the NFW fits that fail do so in a systematic way: the inner
velocity is overestimated, then the model drops below the observed
velocities in the middle and in the outer parts it once again
overpredicts the velocity.  The NFW residuals are most pronounced when
plotted against $R/R_{\rm max}$. The majority of the residuals change
sign at $\sim 0.2R_{\rm max}$ and $\sim 0.7R_{\rm max}$.  As the
radius $R_{\rm max}$ does not have any physical significance, but is
determined by the observations (slit angle, presence of H$\alpha$,
etc.) this indicates that the systematics are due to the choice of
model, rather than being associated with any particular length scale
in the galaxies. Similar conclusions are reached when the residuals
are plotted for the min+gas, constant \MLstar\ and maximum disk cases.

Though this is not readily apparent in Fig.~\ref{nfwiso}, the
residuals for the pseudo-isothermal halo model also show a systematic
behaviour, though at a much lower level than the NFW model.  Here the
residuals do not increase towards the center, and as the typical size
of the residuals is smaller than the uncertainty in the individual
data points, this just shows us that the rotation curve shape is subtly
different from that of a pure pseudo-isothermal halo. This
should come as no surprise given the simplifying assumptions of e.g.\
minimum disk that we have made.

\subsection{The pseudo-isothermal halo}

Of the two models investigated, the pseudo-isothermal halo best
describes the data.  Here we briefly explore some correlations between
the pseudo-isothermal halo model parameters and the parameters
describing the luminous components of the galaxies.  To increase the
range of the parameters we also consider the samples of
\citet{broeils} of (mainly) luminous HSB galaxies, and
\citet{swaters_phd} of late-type dwarf galaxies.  From these samples
we only select bulge-less galaxies brighter than $M_B=-16.5$ to be
consistent with the range of luminosities found in our sample.

\begin{figure} 
\begin{center} 
\epsfxsize=0.8\hsize \epsfbox{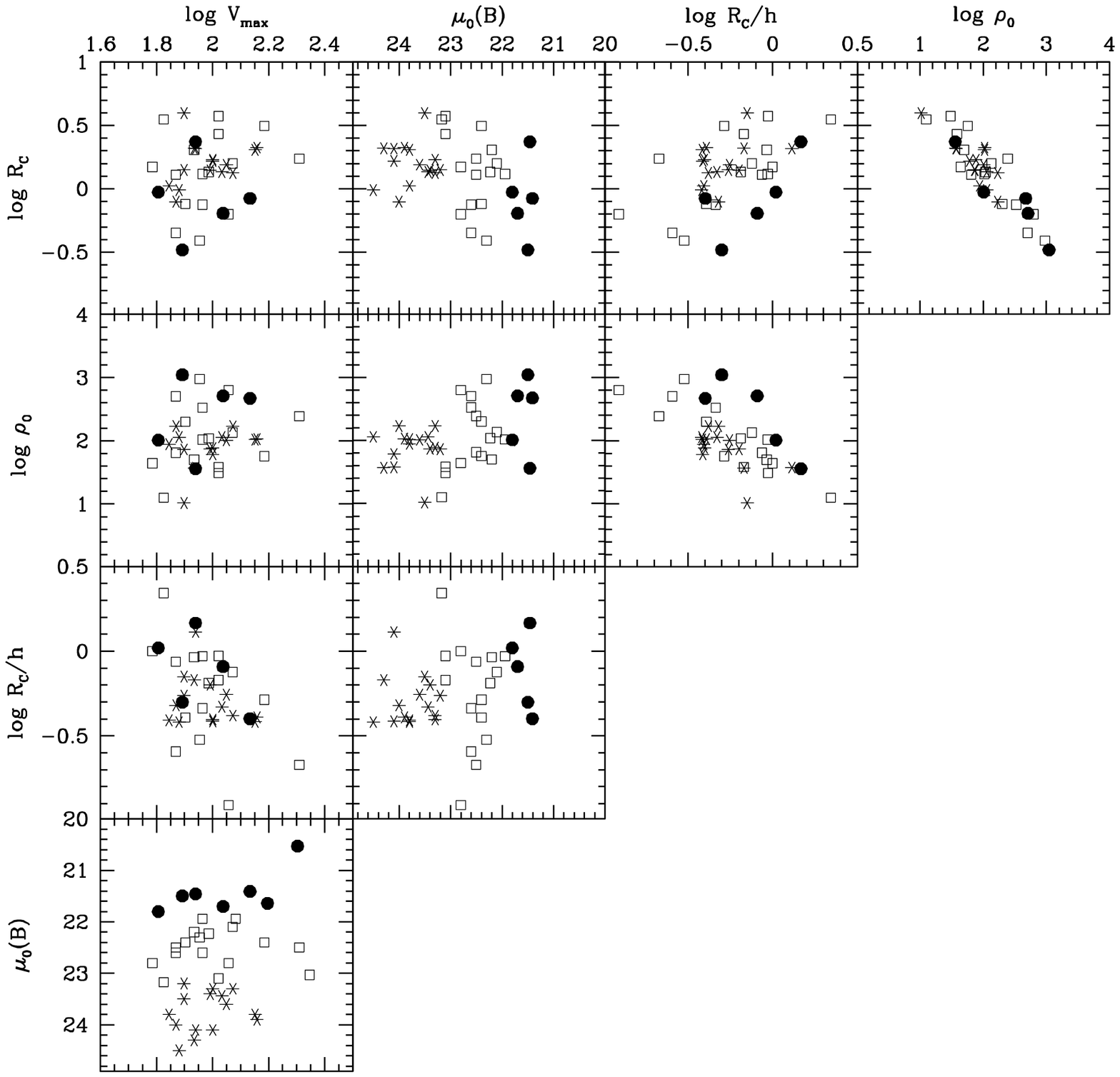}
\figcaption[fig10.ps]{Correlations involving pseudo-isothermal halo parameters 
assuming minimum disk+gas.  Shown are maximum rotation velocity
$V_{\rm max}$ [km s$^{-1}$], surface brightness $\mu_0 $[$B$-mag
arcsec$^{-2}$], central halo density $\rho_0$ [$10^{-3}\ M_{\odot}$
pc$^{-3}$], halo core radius $R_C$ [kpc], and the ratio of core radius
and optical disk scale length $R_C/h$. Included are the bulge-less HSB
galaxies brighter than $M_B = -16.5$ from Broeils (1992) and the
high-quality ($q = 0$ or 1) curves of bright $M_B < -16.5$ dwarfs from
Swaters (1999).  Black circles: $\mu_0(B) < 21.9$ mag arcsec$^{-2}$;
open squares: $21.9 \leq \mu_0(B)
\leq 23.2$; stars: $\mu_0(B) > 23.2$ mag arcsec$^{-2}$. See text for details.
\label{isocor}
}
\end{center}
\end{figure}

Fig.~\ref{isocor} presents the results for the three samples. We show
the min+gas case, which the two comparison samples refer to as their
``minimum disk''.  The most obvious correlation visible in
Fig.~\ref{isocor} is that between $R_C$ and $\rho_0$. This is a
reflection of the fact that these two are correlated through the
asymptotic velocity $V_{\infty}$ (See Eq.~3). Lines of constant
$V_{\infty}$ have a slope of $-\frac{1}{2}$ in the $R_C-\rho_0$
diagram, and the diagram therefore just reflects the limited range in
$V_{\rm max}$ in the samples.  There is an indication that that $R_C$
increases towards lower surface brightnesses, and that $\rho_0$
decreases (as one would expect if LSB galaxies inhabited lower density
halos).

The large scatter in these figures sheds little light on galaxy
formation or the details of pseudo-isothermal halos. How the observed
regularities of galaxy kinematics (like the Tully-Fisher relation) can
emerge from this scatter remains a mystery.

A further analysis is presented in
\citet{paper3} where the mass
density distributions that give rise to the observed rotation curves
are presented. They show that the minimum disk mass density
distributions at small radii can be parametrized by a power law $\rho
\sim r^{\alpha}$ where for the LSB galaxies $\alpha = -0.2 \pm 0.2$,
clearly different from $-1.5 \leq \alpha \leq -1$ as predicted by CDM.
These minimum disk slopes are upper limits.  When stars are properly
taken into account, assuming some reasonable value for $\MLstar$, the
slopes decrease, and become even more consistent with constant density
cores.  Succesfull theories of galaxy formation and evolution that
attempt to model LSB galaxies should thus be able to produce halos
dominated by constant-density cores.

\subsection{The NFW halo}

As noted earlier, the $c$ and $V_{200}$ halo parameters are related.
Here we compare the derived $c$ and $V_{200}$ values with those
predicted by $\Lambda$CDM with the \citet{NFW97} prescription for
$\Omega_m=0.3$, $\Omega_{\Lambda} = 0.7$, and $h = 0.65$ with a COBE
normalized power spectrum.  The values of $c$ depend on the assumed
$\MLstar$, which, as discussed before, is uncertain. Minimum disk
however gives strong upper limits on the values of $c$: when $\MLstar$
is increased the halo needs to compensate by becoming less
concentrated \citep{NFW97}.  Minimum disk models with $c$ values
higher than found in simulations can usually be reconciled with these
observations by increasing $\MLstar$ or introducing a bulge to bring
the $c$-values down, as one can see from the progressive decrease in
$c$ values from minimum disk to maximum disk.

Explaining minimum disk models with concentrations lower than the
simulated values is more difficult.  It indicates one or more of three
problems: failure of the model, failure of the assumption of circular
motion in deriving rotation curves, or a dramatic (non-cosmological)
re-distribution of DM. This last option is not really
understood, and potentially removes any of the predictive power that
the CDM theory has. We will not discuss it here, except to note that
the most plausible effect, adiabatic contraction, further concentrates
the DM, making the problem worse.

As an aside we note here that the min+gas case sometimes gives
slightly higher $c$-values than the simple minimum disk case. In most
cases this is due to a central depression in the
\HI surface density that gives rises to imaginary rotation
velocities, which have to be compensated for by the halo. Also some of
the outer rotation velocity is explained by the gas rotation curve,
yielding a halo curve that bends more at small radii. Consequently the
halo model tends to be slightly more concentrated.

\begin{figure} 
\begin{center} 
\epsfxsize=0.8\hsize \epsfbox{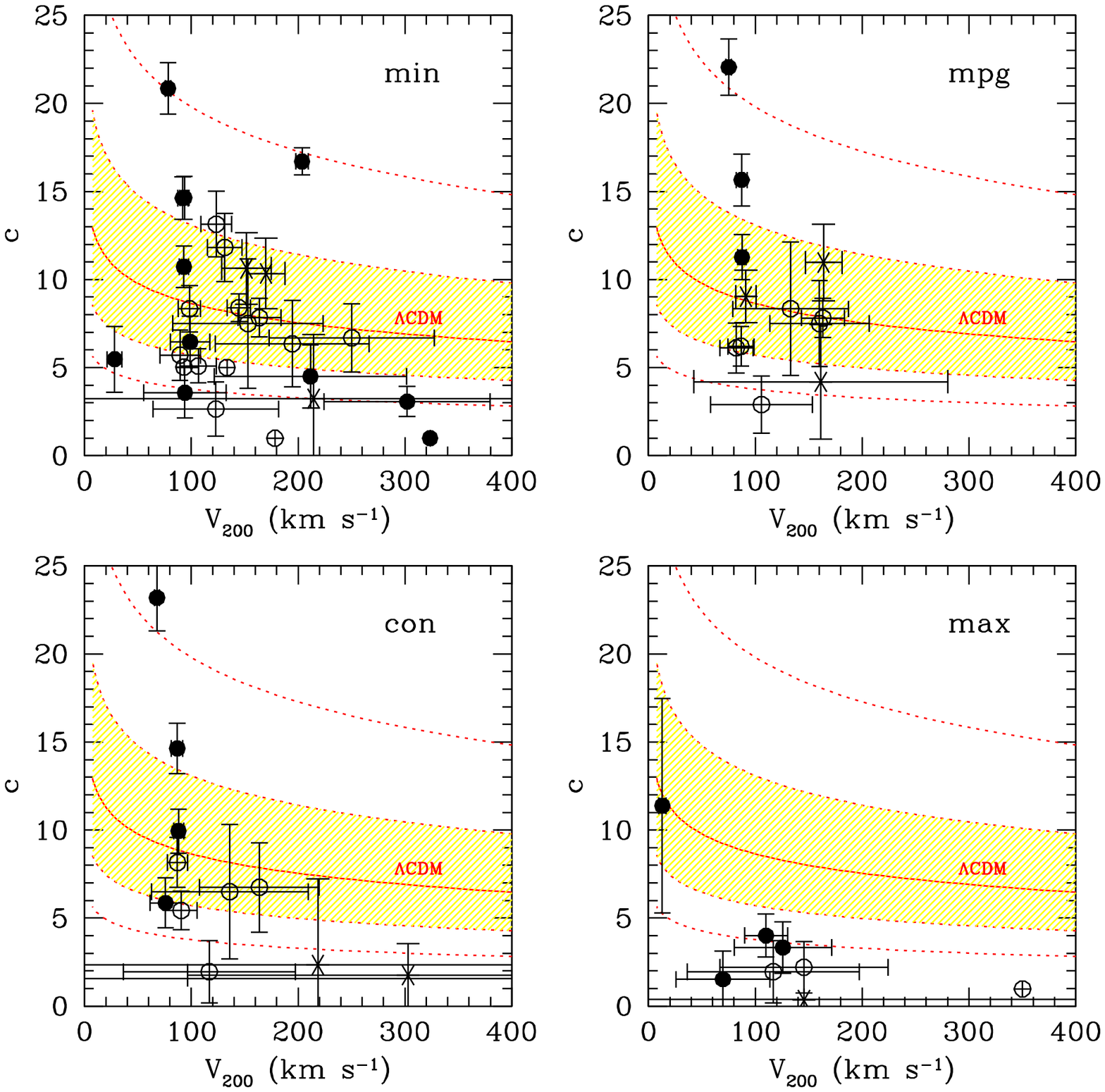}
\figcaption[fig11.ps]{The NFW halo concentration parameter $c$ plotted
against the halo rotation velocity $V_{200}$ for the four different
\MLstar\ cases discussed in this paper. Black dots represent good fits
$( p > 0.95)$, open dots average quality fits $(0.05 < p <
0.95)$. Crosses represent bad fits ($p<0.05$). Good fits are primarily
found at $V_{200} < 100$ km s$^{-1}$. Maximum disk is clearly
inconsistent with NFW.  The line labeled ``$\Lambda$CDM'' shows the
prediction for that cosmology derived from numerical models.  The grey
area encloses the 1$\sigma$ uncertainty \citep{bul99}.  The upper and
lower dotted line show the 2$\sigma$ uncertainty.  The minimum disk
panel shows both Samples I and II. The other three panels only show
Sample I.
\label{csumm}}
\end{center}
\end{figure}

Fig.~\ref{csumm} shows the derived $c$ and $V_{200}$ values and
compares them with the $\Lambda$CDM predictions.  For the minimum disk
case we show both Samples I and II, for the other three $\MLstar$
values only Sample I is shown.  The data points are coded to indicate
their significance level $p$.

Several points can be made about Fig.\ref{csumm}. Firstly, the
bottom-right plot clearly shows that maximum disk is inconsistent with
the NFW halos expected to arise in $\Lambda$CDM.  Secondly, there is
some correlation between the significance of the fits and their
position in the $c-V_{200}$ diagram.  In the minimum disk case the
majority (11 out of 14) of the $p>0.95$ points are found at $V_{200}
\lesssim 100$ km s$^{-1}$. Most (17 out of 19) of the $p<0.95$ points
are found to the right of this line. This division becomes more clear
in the minimum disk+gas and constant \MLstar\ plots.  As the high
$V_{200}$ values tend to occur at lower $c$, this is likely to be the
effect of a NFW halo trying to fit a solid-body-like curve, by hiding
its curvature outside the visible galaxy, i.e., by decreasing $c$ and
increasing $V_{200}$.

\begin{figure} 
\begin{center} 
\epsfxsize=0.8\hsize \epsfbox{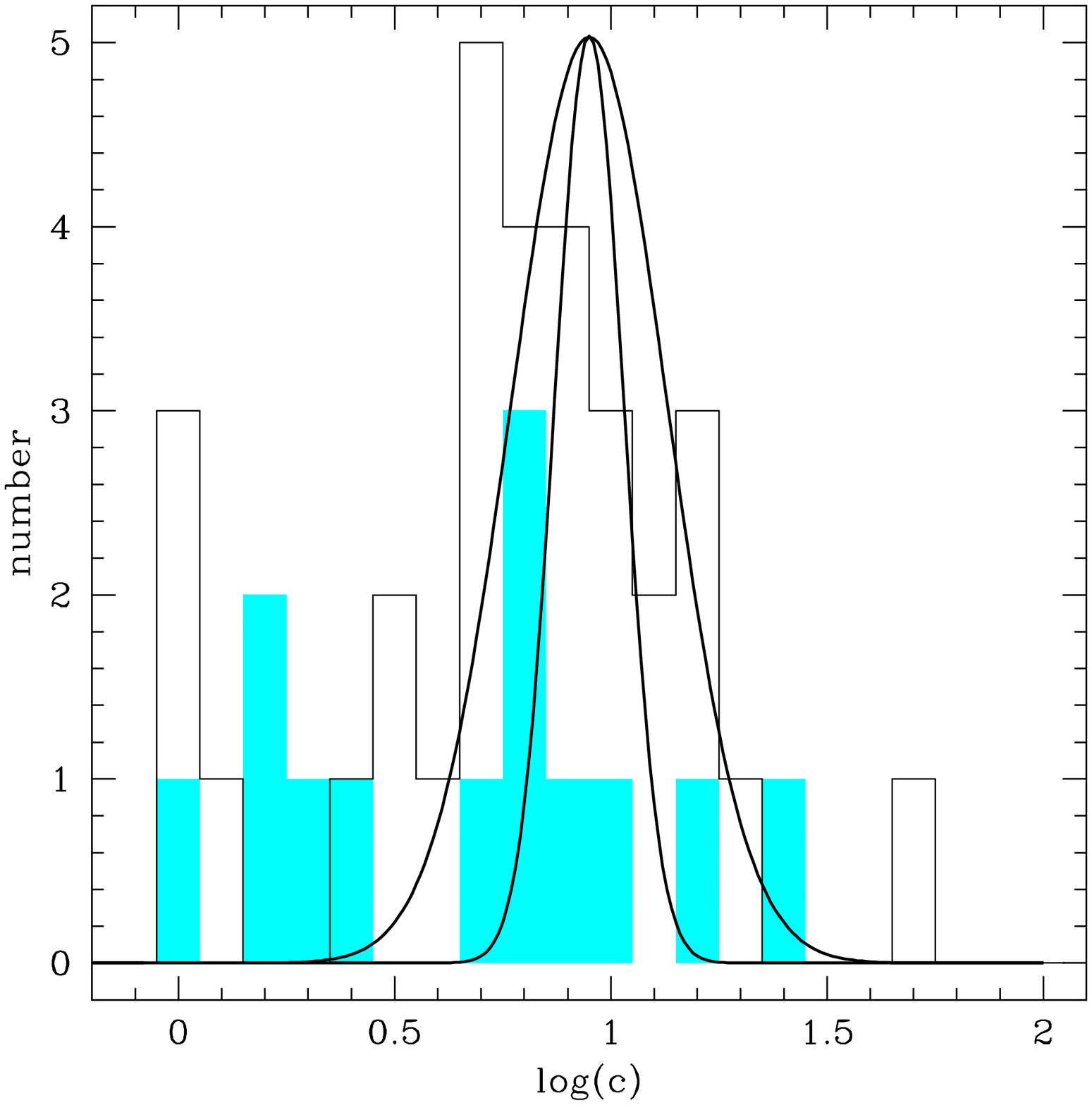}
\figcaption[fig12.ps]{Distribution of the $c$-parameter for the minimum 
disk case (open histogram) and the constant \MLstar\ case (solid
histogram). Over-plotted is the theoretical log-normal distribution
for a $\Lambda$CDM cosmology, derived from independent numerical
simulations by \citet{jing99} and \citet{bul99}. The former finds a
log-normal distribution with a logarithmic dispersion $\sigma_{c}
=0.08$. The latter finds a wider log-normal distribution with
$\sigma_{c} =0.18$.  The observed low-$c$ tail is not consistent with
either theoretical distribution. The theoretical distributions have
been arbitrarily normalized to coincide with the maximum of the
observed distribution.
\label{cdist}}

\end{center}
\end{figure}

Thirdly, the distribution of points does not agree with that predicted
by the numerical models \citep{jing99,bul99}. There are more points
above, but more importantly, below the 1$\sigma$ lines than expected.
This low-$c$ tail consists of fits that have a high to reasonable
significance $p$ associated with them.  We show the distribution again
in Fig.~\ref{cdist}. The two histograms are for minimum disk (open
histogram, Samples I and II) and constant \MLstar\ (filled histogram,
Sample I). Over-plotted are log-normal distributions showing the
distribution derived from numerical simulations. Unfortunately, these
simulations do not agree on the value of the dispersion.  The
$\Lambda$CDM model by
\citet{bul99} gives a logarithmic dispersion $\sigma_{c} \simeq 0.18$,
while the distribution for relaxed $\Lambda$CDM halos as found by
\citet{jing99} has a logarithmic dispersion of $\sigma_{c} \simeq 0.08$.  
The observed distribution is clearly wider than either theoretical
one. By changing the cosmology of the model one can change the mean of
the distribution (e.g.\ OCDM has a mean $\log c = 1.25$)
\citep{jing99}, but the width hardly changes. Thus one can possibly
shift the model to higher $c$ to fit the high-$c$ end of the
distribution, but it is impossible to explain the large observed
low-$c$ tail with the kind of log-normal distribution one derives from
the simulations. 

\subsection{Morphology}

Rotation curves have the implicit assumption of circular motion.  Can
non-circular motions affect the rotation curves? As the NFW models
show the largest residuals in the centers of some of the LSB galaxies,
it is possible that they could be affected by non-circular motions due
to non-axisymmetric components. We will investigate the matter here by
comparing the morphology of our galaxies with the quality of the fits.

Table~\ref{morphtable} contains a short description of the morphology
of the galaxies, where we have focussed on the central parts. In the
Table ``core'' refers to a galaxy whose central light distribution can
best be described by an axisymmetric model, presumably implying
negligible non-circular motions in the inner part. The word ``core''
is used very loosely here. It does not necessarily indicate the
presence of a bulge or massive central component, but is just an
indication of the (deprojected) round shape of the isophotes in the
inner part of the galaxy.  ``Bar'' indicates a central morphology
dominated by a bar-like structure, usually magellanic, that may
indicate the presence of non-circular motions.

The results are summarized in Table~\ref{morphstattable} (for the
minimum disk assumption). The conclusion is that there is no clear
dependence of residual velocity on morphology. There is thus no
indication that the failure of NFW to fit some galaxy rotation curves
can be attributed to the presence of bars or non-circular motions.

\section{The maximum disk}

As noted in the Introduction, the inner rotation curves of HSB
galaxies can usually be well explained by scaling up the rotation
curve derived from the light distribution.  This maximum disk
procedure results in \MLstar\ values that are reasonably consistent
with those derived from stellar population synthesis models
\citep{marc_phd97, palunas_maxdisk,maxdisk86}.  Furthermore, bars seem
to demand near-maximal disks in HSB galaxies
\citep{victor_maxdisk,weiner00}

The matter of maximum disk in LSB galaxies was first discussed by
\citet{edb_rot} who noted that from a stellar population 
point of view maximum disk demanded unreasonably high $\MLstar$
values. Substantial amounts of DM were still needed within
the optical disk to explain the observed \HI curves.  SMT revisited
the subject and noted that the slightly steeper slopes they found
using their H$\alpha$ curves, enabled them to scale up the disk
rotation curve by an even larger factor.  The maximum disk \MLstar\
value is extremely sensitive to the inner slope, and only a very small
increase is needed to change
it by a significant factor.  In LSB galaxies, the maximum disk
\MLstar\ values are thus much larger than expected on the basis
of colors, metallicities and star formation histories.  The higher
\MLstar\ values as found by SMT worsened this problem (a consequence  already
noted in \citealt{edb_rot}), despite the fact that their maximum disk
models could slightly better reproduce the observed inner curve.

Though it seems unlikely that the maximum disk results can be
explained using ``reasonable'' stellar populations, given what we know
about the star formation history, dust content and metallicity of LSB
galaxies, the matter is still relevant for exploring possible baryonic
disk DM scenarios.  To explain maximum disk in LSB galaxies
purely in terms of baryons one has to assume a large amount of unseen
material in the form of e.g.\ cold molecular gas, optically thick
neutral hydrogen, low mass stars or non-standard initial mass
functions.  It should be noted though that many of these hypothetical
mass components would however violate constraints imposed by disk
stability \citep{athan87, mihos_lsb}, near-infra red colors
\citep{bell_lsb00, bell_pops}, and would possibly introduce a surface
brightness segregation in the baryonic TF-relation
\citep{mcg_barytf}

\begin{figure} 
\begin{center} 
\epsfxsize=0.8\hsize \epsfbox{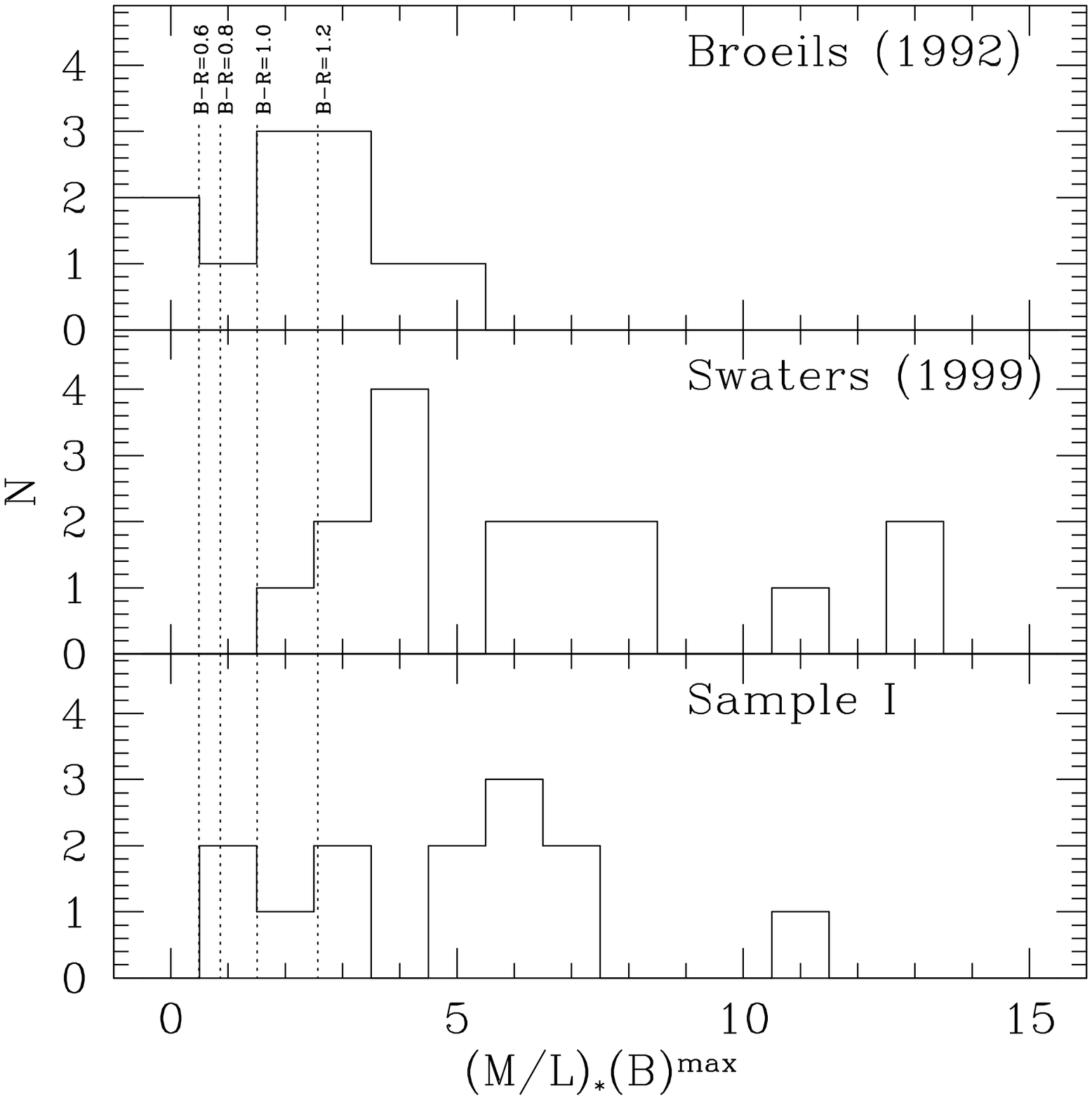}
\figcaption[fig13.ps]{Histograms of the maximum disk \MLstar\ values. The top panel
refers to bulge-less galaxies from the collection of Broeils (1992)
brighter than $M_B = -16.5$. The middle panel refers to dwarf galaxies
brighter than $M_B = -16.5$ from Swaters (1999) with quality index 0
or 1 (very good to good). The bottom panel shows Sample I. Also
indicated are the values for \MLstar\ using population synthesis
models by Bell \& de Jong (2001) assuming a simple Salpeter IMF.
\label{mlhisto}}

\end{center}
\end{figure}

In Fig.~\ref{mlhisto} we compare the maximum disk $B$-band \MLstar\
ratios\footnote{These were derived by converting our $R$-band values
using the $B-R$ color. We converted our data to $B$-band, rather than
converting the HSB data to $R$-band, as the colors of LSB galaxies are
better determined than those of the Broeils HSB galaxies. The color
gradients in LSB galaxies are small, so systematic effects are
negligible.}  for Sample I, with those derived by
\citet{broeils} for a sample of mostly HSB galaxies and by  
\citet{swaters_phd} for a sample of dwarfs.  We again show 
only bulge-less galaxies brighter than $M_B = -16.5$.  Also indicated
are \MLstar\ values from \citet{bell_tfproc}, who tabulate stellar
mass-to-light ratios for various star formation histories and
population synthesis models as a function of color. Here we show
representative values (assuming a simple Salpeter IMF) spanning the
color range exhibited by late-type HSB galaxies, gas-rich dwarfs and
LSB galaxies.

The \MLstar\ values for HSB galaxies agree to within a factor of 3 and
can be considered to be (close to) maximum disk.  The values found for
LSB galaxies and dwarfs are less easily reconciled with the model
values. Observationally values up to $\MLstar(B) = 15$ are found,
while the typical model value (again for a simple Salpeter IMF) is
$\MLstar(B) \simeq 0.9$ for $B-R=0.8$ (the average color for a
dwarf/LSB). This discrepancy cannot be explained with extinction or
population effects.  Extinction in dwarfs and LSB galaxies is less
than in HSB galaxies \citep{marc_bimodal}, and a factor of $\sim 17$
(3.0 mag) extinction is hard to reconcile with the known properties of
LSB galaxies. Line-of-sight extinctions observed towards H{\sc ii}
regions in LSB galaxies are never as large
\citep{mcg_abund94,edb_abund98}.

Apart from changing the IMF in an \emph{ad hoc} way it is
hard to see how such high \MLstar\ values can be reached given the
constraints imposed by what we know about the star formation history
(low SFR in past and at present) and the blue (optical and NIR) colors
of LSB galaxies \citep{edb_phot95, mcg_opt94, vdhoek00, bell_lsb00,
bell_pops, dejong_grad}. It is likely that the maximum disk values as
found in LSB galaxies are not representative of the evolutionary stage
of these galaxies.  While the maximum disk prescription now has
somewhat greater success in predicting the inner shape of the rotation
curves of LSB galaxies, it requires stellar mass-to-light ratios which
are too large for the stellar populations in these
galaxies. The mass discrepancies are still large; all this does is
move the DM from halo to disk.

\subsection{The maximum surface density of a disk}

Just as the minimum disk assumption gives us an upper
limit on the amount of DM implied by rotation curves, the
maximum disk hypothesis gives us an upper limit on the amount of mass
that could potentially be hidden in a disk. It is therefore still
useful to ask ourselves what these maximum disk upper limits imply for
the stellar disks.

\begin{figure} 
\begin{center} 
\epsfxsize=0.8\hsize \epsfbox{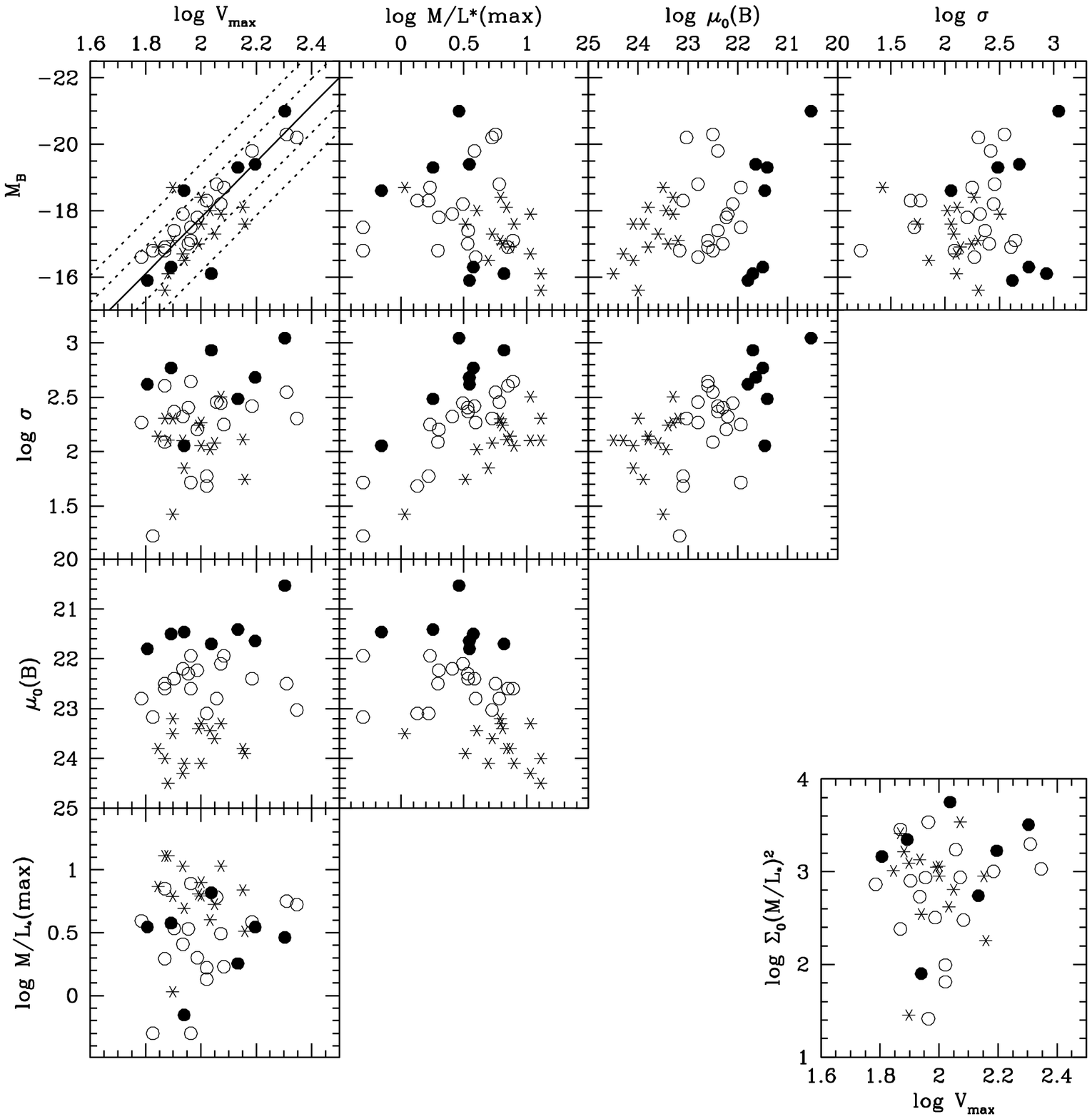}
\figcaption[fig14.ps]{Correlations between maximum disk surface density $\sigma $[$M_{\odot}$ pc$^{-2}$], 
surface brightness $\mu_0 $[$B$-mag arcsec$^{-2}$], maximum disk
\MLstar(B) [$M_{\odot}/L_{\odot,B}$], maximum observed rotation
velocity $V_{\rm max}$ [km s$^{-1}$] and luminosity $M_B$ [mag].  Included
are again the bulge-less bright HSB galaxies from Broeils (1992) and
the high-quality curves of bright dwarfs from Swaters (1999), as
described in the previous figure caption.  Black circles: $\mu_0(B) <
21.9$ mag arcsec$^{-2}$; open circles: $21.9 \leq \mu_0(B)
\leq 23.2$; stars: $\mu_0(B) > 23.2$ mag arcsec$^{-2}$.
The inset panel in the lower-right corner shows the product
$\Sigma_0(\Upsilon_*)^2$, where $\Sigma_0$ is the central surface
brightness expressed in $L_{\odot}$ pc$^{-2}$. This product should be
constant with small scatter for a maximum disk interpretation of the
TF relation. The TF relation is shown in the top-left corner.  It has
a slope of $-8.4$.  The dotted lines represent the $1\sigma$ and
$2\sigma$ scatter where $\sigma = 0.81$ mag.  This scatter is reduced
to 0.51 mag when the 4 most outlying points are omitted (all galaxies
with low inclinations).
\label{maxdisk}}

\end{center}
\end{figure}

Maximum disk means maximum surface density (luminous surface density
times \MLstar), and therefore gives an absolute upper limit on the
mass surface density in stellar disks (for mass components that are
distributed like the stars).  Figure~\ref{maxdisk} summarizes the
maximum disk results for the Sample I LSB galaxies as well as the
\citet{broeils} and \citet{swaters_phd} HSB and dwarf samples.
We plot the maximum disk \MLstar\ values, as well as the luminosity,
rotation velocity, surface brightness, and maximum disk surface
density.  The data are divided into three surface brightness bins. As
already shown in \citet{edb_rot}, at fixed $V_{\rm max}$ LSB galaxies
have higher maximum \MLstar\ values than HSB galaxies.

Figure~\ref{maxdisk} also shows the maximum surface density $\sigma$.
As the decrease in surface brightness is faster than the increase in
maximum disk $\MLstar$, towards low surface brightnesses the maximum
surface density $\sigma$ in a disk decreases with surface
brightness. The $\mu_0 - \sigma$ panel suggests that there is a
well-defined upper limit to the maximum surface density disks can
attain. Even under maximum disk, LSB galaxy disks have on average
lower surface densities than HSB galaxy disks, again putting limits on
the amount of baryonic mass one can hide in these disk.

Figure~\ref{nfwiso} shows that even in the maximum disk case most LSB
galaxies have $V_{\rm max}$(disk) $< V_{\rm max}$(observed). Therefore, even
in the maximum disk case a moderate amount of DM is still
required in the optical disk.  It is therefore hard to explain the TF
relation for LSB galaxies in the context of maximum disk: the stellar
disk then needs to provide the luminosity, \emph{and} the necessary
rotation velocity. LSB galaxies would deviate systematically from the
HSB TF relation, which is evidently not the case
\citep{zwaan_tf95}. 

This is illustrated in the inset panel in Fig.~\ref{maxdisk}
(lower-right). Using the arguments in
\citet{zwaan_tf95} we find that $\Sigma_0(\Upsilon)^2$ needs to be
constant for galaxies to obey a TF relation independent of surface
brightness. If all galaxies were truly maximum disk (in the sense that
$V_{\rm max}$(disk) $\simeq V_{\rm max}$(observed), one could replace this by
the requirement that $\Sigma_0(\Upsilon_\star)_{\rm max}^2$ needs to be
constant. The lower-right panel shows that this is not the case: at
fixed $V_{\rm max}$ there is a substantial scatter which would translate
in $\sim 5$ mag scatter in TF.  Clearly the observed scatter is much
smaller, and this shows the clear need for an additional mass
component to make TF work. In other words, maximum disk for \emph{all}
galaxies and TF are incompatible.

\section{Conclusions}

The most important conclusion from this work is that the large
majority of the high-resolution rotation curves presented here 
prefer the pseudo-isothermal core-dominated halo model.  For a small
number of galaxies neither the pseudo-isothermal nor the NFW models
are an adequate description of the data.  This should not come as a
surprise as the true DM distribution is likely to be more complex than
the models presented here. Nevertheless, the general trend is that for
almost all galaxies discussed here the relative quality of the fits
using the pseudo-isothermal model is better than those for the NFW
model.

For a small numer of galaxies the NFW model provides a good fit, but
generally the concentrations derived from the observed rotation curves
are lower than predicted by the simulations.  This is hard to fix: the
most likely effect which may alter the initial cosmological NFW halo
is adiabatic contraction, but this has the effect of making the final
(observed) halo \emph{more\/} concentrated, so one would have to start
off with (cosmologically relevant) halos that are even
\emph{less} concentrated. 

\begin{figure} 
\begin{center} 
\epsfxsize=0.8\hsize \epsfbox{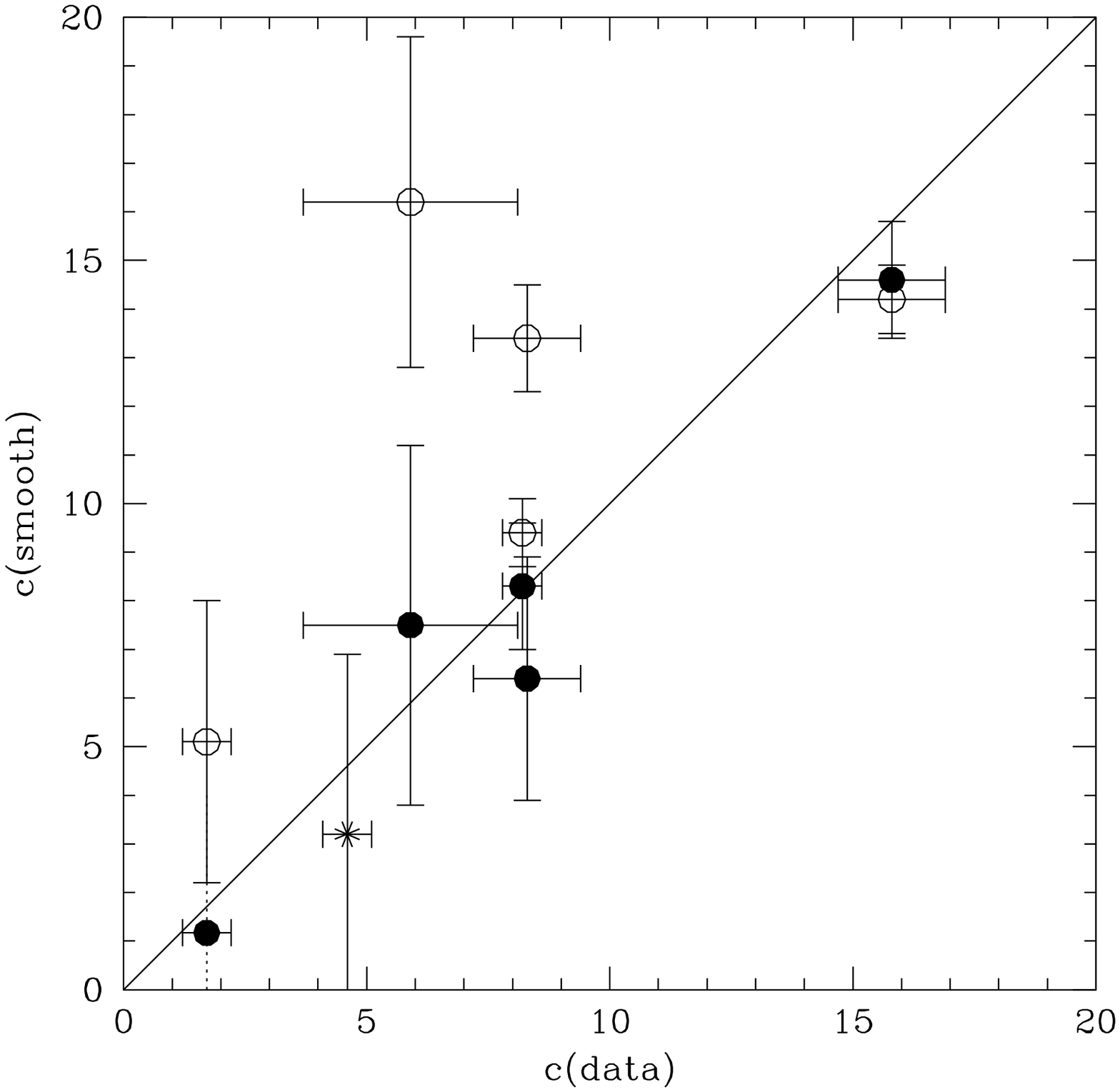}
\figcaption[fig7.ps]{Comparison of the NFW halo parameter $c$ fitted to 
the SMT data.  The horizontal axis represents $c$-values derived from
the raw data; the vertical axis those derived from the smooth curves.
The open symbols compare the values derived from the raw data with
those derived from the SMT smooth curves. The filled symbols compare
the raw data results with the values derived from our smooth curves.
The star symbol represents the result from our analysis of our data
for F568-3.  The dotted line in the lower-left corner indicates that
no realistic errorbars could be derived for this fit. Our analysis of
the raw, unsmoothed data, of our smoothed versions of these curves,
and of the SMT smooth curves show good agreement.  The only exceptions
are F563-V2 and F568-1, for which the SMT curves give larger
concentrations.  While the formal fits differ significantly in these
cases, the differences in the curves being fit are subtle (see
Fig.~\ref{swcomp}).  This illustrates the importance of high accuracy
data.
\label{nfwswaters}}
\end{center}
\end{figure}

It is worrying that for one or two extreme cases the difference
between ``CDM does work'' or ``CDM doesn't work'' depends on subtle
differences in data, data handling, or
analysis. Figure~\ref{nfwswaters} illustrates that opposite claims can
sometimes be made from the same data. Hence we re-iterate the need for
the highest quality data of a large sample, in order to minimize these
effects.

We refer to \citet{paper3} where it is shown that {\it all\/} data
presented here are consistent with a core-dominated model; the good
NFW fits that are found for a number of LSB galaxies can be attributed
to resolution effects.

We summarize our results as follows.

\begin{itemize}

\item Pseudo-Isothermal halos are a better description of the data than NFW halos.

\item The number of galaxies that cannot be fit with NFW halos is 
significantly larger than the number of galaxies that cannot be fitted
with the pseudo-isothermal model.

\item The quality of the fit is not obviously related to morphology, 
luminosity, or surface brightness.
\item A larger number of low-$c$ NFW halos is found than one would expect based on the distribution derived from CDM simulations.
\item If one were to construct models that would have the correct
$c$-$V_{200}$ values as predicted by cosmology, then the resulting
\MLstar\ values would be too low to be consistent with stellar
population numbers. The shape of the curves would still be wrong.
\item The maximum disk prescription works to predict the inner rotation 
curve shape to some extent, but gives mass-to-light ratios which are
too high to be accounted for by stellar population synthesis models.
\item Applying the maximum disk  values yields absolute upper limits on 
the disk mass surface density that is strongly correlated with surface
brightness.

\end{itemize}

\acknowledgements

We thank Roelof Bottema and Rob Swaters for their helpful comments on
early drafts of this paper. We thank the anonymous referee for a
thorough examination of the data. The work of SSM is supported in part
by NSF grant AST9901663.  This research has made use of the NASA/IPAC
Extragalactic Database (NED) which is operated by the Jet Propulsion
Laboratory, California Institute of Technology, under contract with
the National Aeronautics and Space Administration.  This research has
made use of NASA's Astrophysics Data System Abstract Service.

\clearpage

\begin{figure} 
\figurenum{1}
\begin{center} 
\epsfxsize=0.95\hsize
\epsfbox{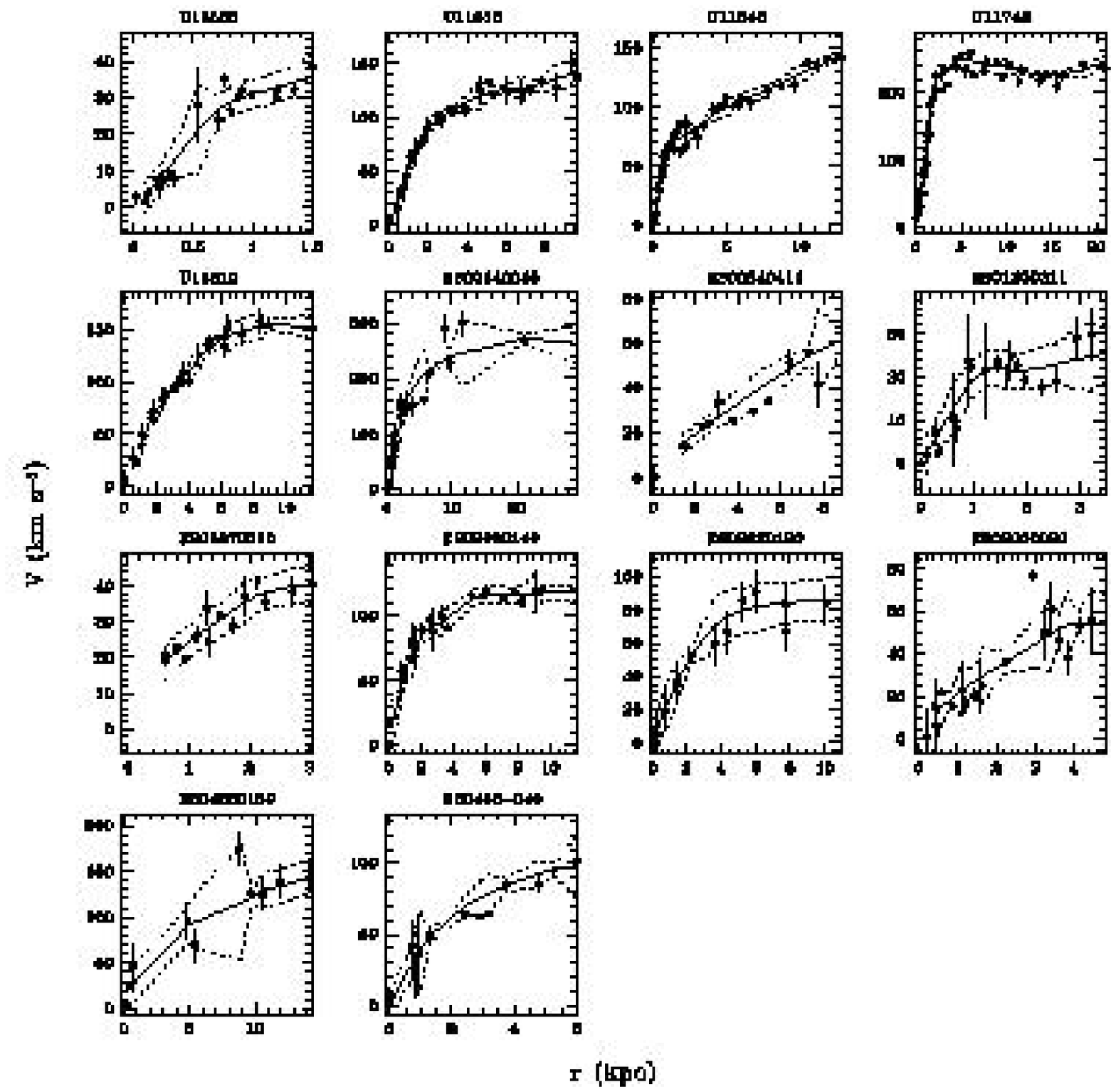}
\figcaption[fig1b]{{\sc continued}}
\end{center}
\end{figure} 

\begin{figure} 
\begin{center} 
\figcaption{Mass models assuming NFW halo (left) and
pseudo-isothermal halo (right). Within each panel the left column shows the
best fitting model, the right column shows the probability
distribution of the halo parameters. For a full description see text
(Sect.~4). {\bf Figure 3 not available due to size limitations. All figures
available at http://www.atnf.csiro.au/$\sim$edeblok/papers/deblok.ps.gz}
\label{nfwiso}}
\end{center}
\end{figure} 

\clearpage 

\begin{figure} 
\begin{center} 
\epsfxsize=0.8\hsize \epsfbox{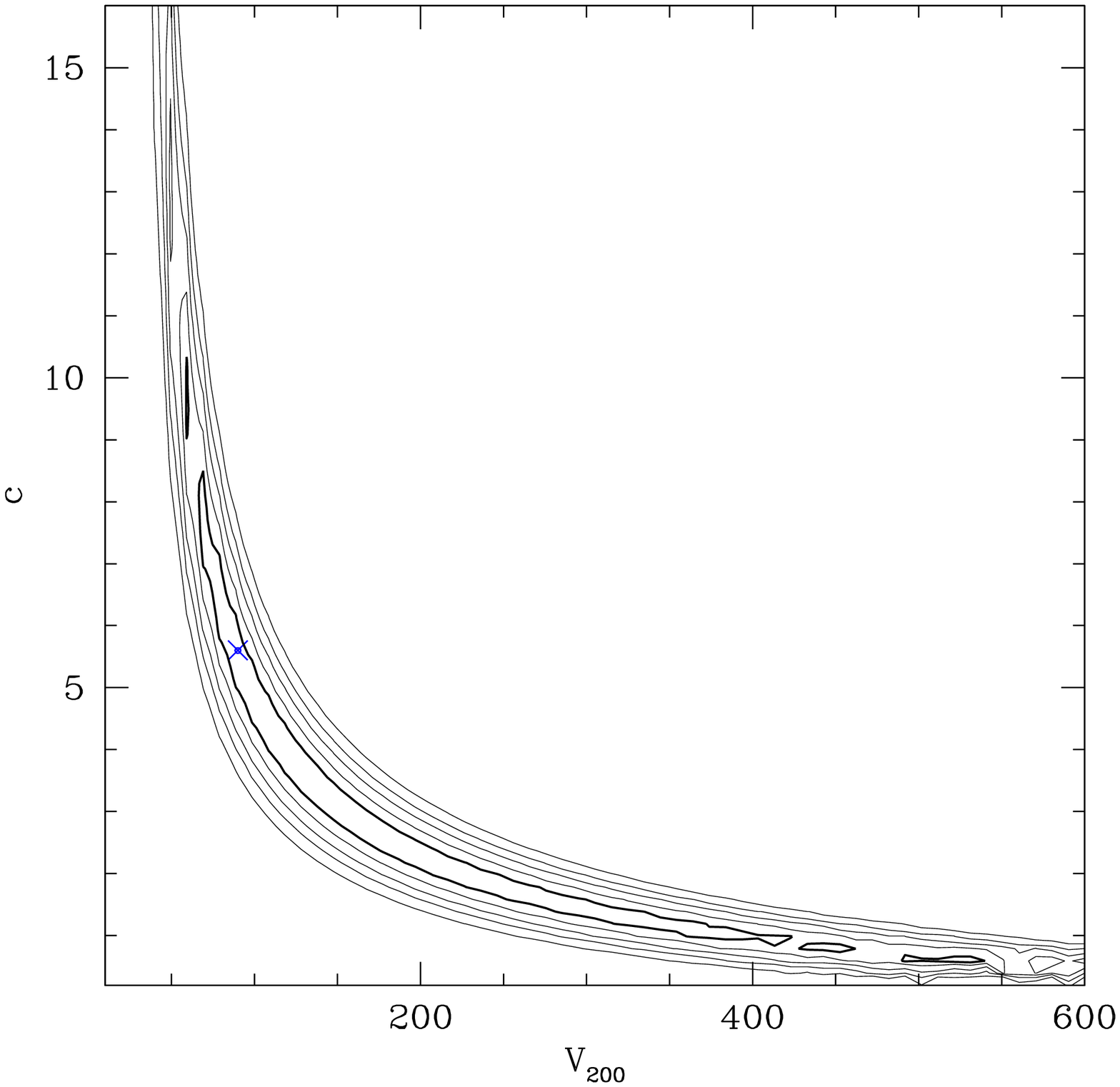} 
\figcaption[fig4.ps]{The error contours for the minimum disk NFW fit
of F583-4 drawn in linear $c-V_{200}$ space. Contour values are as in
Fig.~\ref{nfwiso}.\label{linchi2}}
\end{center}
\end{figure}

\clearpage

\begin{deluxetable}{lrcccccccc}
\tablecaption{Sample I: galaxies with photometry\label{fgalsample}}
\tablewidth{0pt}
\tablehead{
\noalign{\vskip 2pt}
\colhead{Name}& \colhead{$D$} & \colhead{$\mu_0(B)$}&\colhead{$h$}&\colhead{$M_{\rm abs}(B)$}&\colhead{$R_{\rm max}$}&\colhead{$V_{\rm max}$} &\colhead{$V_{\rm hel}$}&\colhead{$i$}\\
\colhead{}& \colhead{(Mpc)}&\colhead{(mag/$\sq "$)}&\colhead{(kpc)}&\colhead{(mag)}&\colhead{(kpc)}&\colhead{(km s${^{-1}}$)} &\colhead{(km s${^{-1}}$)}&\colhead{$(^{\circ})$}\\
\colhead{(1)}& \colhead{(2)}& \colhead{(3)}& \colhead{(4)}& \colhead{(5)}& \colhead{(6)} & \colhead{(7)}& \colhead{(8)}& \colhead{(9)}
}
\startdata
F563-1 & 45 & 23.6 & 2.8 & -17.3 & 17.7 & 112 & 3502 & 25 \\ 
F563-V2 & 61 & 22.1 & 2.1 & -18.2 & 9.2 & 118 & 4312 & 29\\ 
F568-1 & 85& 23.8 & 5.3 & -18.1 & 14.9 & 142 & 6524 & 26\\ 
F568-3 & 77 & 23.1 & 4.0 &-18.3 & 16.5 & 105 & 5913 & 40\\ 
F568-V1 & 80 & 23.3 & 3.2 & -17.9& 19.0 & 118 &5768 & 40\\ 
F571-8 & 48 & 23.9\tablenotemark{a} & 5.2& -17.6\tablenotemark{a} & 15.6 & 144 & 3768 & 90\\ 
F574-1 & 96 & 23.3\tablenotemark{a} & 4.3 &  -18.4\tablenotemark{a} & 15.4 & 100& 6889 & 65\\ 
F579-V1 & 85 & 22.8\tablenotemark{a} & 5.1 &  -18.8\tablenotemark{a} & 17.3 & 114& 6305 & 26\\ 
F583-1 & 32 & 24.1 & 1.6 &  -16.5 & 14.6 & 87 & 2264 & 63\\
F583-4 & 49 & 23.8\tablenotemark{a} & 2.7 &   -16.9\tablenotemark{a} & 10.0 & 70 &3617 & 55\\ 
U5750 & 56 & 23.5\tablenotemark{a} & 5.6 &  -18.7\tablenotemark{a} & 21.8 & 79 &4177 & 64\\ 
U6614 & 85 &
  23.4 & 8.1 &  -20.3 &   62& 204 & 6371 & 36 
\enddata
\tablecomments{(2) Distance computed assuming Hubble flow after correction for
galactic rotation and Virgocentric flow. (6) Maximum radius of rotation curve. (7) Maximum velocity in rotation curve. (8) Heliocentric systemic velocity.
Photometric and distance data from de Blok \& McGaugh (1997). $R_{\rm max}$, $V_{\rm max}$ and $V_{\rm sys}$ derived from new optical curves.}
\tablenotetext{a}{Converted from $R$-band assuming $B-R = 0.9$.}
\end{deluxetable}

\begin{deluxetable}{lrrrrrr}
\tablecaption{Sample II: galaxies without photometry\label{esosample}}
\tablewidth{0pt}
\tablehead{
\noalign{\vskip 2pt}
\colhead{Name}& \colhead{$D$} & \colhead{$V_{\rm hel}$}&\colhead{$M_{\rm abs}(B)$}&\colhead{$R_{\rm max}$}&\colhead{$V_{\rm max}$} & \colhead{$i$}\\
\colhead{}& \colhead{(Mpc)}&\colhead{(km s${^{-1}}$)}&\colhead{(mag)}&\colhead{(kpc)}&\colhead{(km s${^{-1}}$)}& \colhead{$(^{\circ})$}\\
\colhead{(1)}& \colhead{(2)}& \colhead{(3)}& \colhead{(4)}& \colhead{(5)}& \colhead{(6)} & \colhead{(7)}}
\startdata
 F730-V1	&144 & 10714& \nodata &	 11.9 & 145 & 50\\
 U4115		&3.2 & 343  & -12.4 &   1.0    & 40 & 74 \\
 U11454		&91  & 6628 & -18.6\tablenotemark{a}&  11.9 & 152 & 64\\
 U11557		&22  & 1390 & -20.0& 6.2 & 95 & 36\\
 U11583		&5 & 128  & -14.0\tablenotemark{a}& 1.5 & 36 & 83\\
 U11616		&73  & 5244 & -20.3\tablenotemark{a}& 9.6& 143 & 60\\
 U11648		&48  & 3350 & -21.0\tablenotemark{a}& 12.7 &145 & 90\\
 U11748		&73  & 5265 &  -22.9\tablenotemark{a}& 21.0 & 242 & 78\\
 U11819 	&60  & 4261  & -20.3\tablenotemark{a}& 11.7& 153 & 66\\
 E0140040 	&212 & 16064 & -21.6& 29.2& 263 & 35\\
 E0840411 	&80  & 6200  & -18.1& 8.9& 61 & 90\\
 E1200211 	&15  & 1314  & -15.6& 3.5& 25 & 70\\
 E1870510 	&18  & 1410  & -16.5& 3.0& 40 & 58\\
 E2060140 	&60  & 4704  & -19.2& 11.6& 118 & 39\\
 E3020120 	&69  & 5311  & -19.1& 11.0 & 86 & 55\\
 E3050090       &11  & 1019  & -17.3&  4.8 & 54 & 53\\
 E4250180 	&86  & 6637  & -20.5& 14.4& 145 & 33\\
 E4880049 	&22  & 1800  & -16.8 & 6.0& 97 & 63\\
\enddata
\tablecomments{Columns (2) and (3): distance $D$ was calculated from 
$V_{\rm hel}$ after correcting for galactic rotation and assuming pure
Hubble flow with $H_0 = 75$ km s$^{-1}$ Mpc$^{-1}$. Column (4):
Absolute magnitude computed using apparent magnitudes from ESO-LV and
RC3, and are corrected for foreground Galactic extinction.}
\tablenotetext{a}{The apparent magnitude is Zwicky magnitude 17, and therefore very uncertain.}
\end{deluxetable}

\begin{deluxetable}{lrrrrrrr}
\renewcommand{\arraystretch}{.6}
\tablecaption{Modeled Rotation Curves \label{electable}}
\tablewidth{0pt}
\tablehead{
\noalign{\vskip 2pt}
\colhead{Galaxy} & \colhead{$R$} & \colhead{$R$} 
 & \colhead{$V_{gas}$\tablenotemark{a,b}}
 & \colhead{$V_{disk}$\tablenotemark{a,c}}
 & \colhead{$V_{bulge}$\tablenotemark{a,c}}
 & \colhead{$V_{obs}$} & \colhead{$\sigma_{V}$} \\
\colhead{} & \colhead{(arcsec)} & \colhead{(kpc)}
 & \colhead{(km/s)} & \colhead{(km/s)}
 & \colhead{(km/s)} & \colhead{(km/s)} & \colhead{(km/s)}
}
\startdata
F583-1 & & & & & & & \\
&    0.3  &   0.1  & -0.1   &  0.4    &  0    &   1.1   &  11.1 \\
&    2.8  &   0.4  & -0.9   &  4.0    &  0    &  10.0   &   7.0 \\
&    5.0  &   0.7  & -1.5   &  6.7    &  0    &  17.4   &   9.6 \\
&    6.9  &   1.0  & -2.2   &  8.5    &  0    &  23.5   &  11.2 \\
&    9.0  &   1.4  & -2.8   & 10.1    &  0    &  31.0   &   5.2 \\
\enddata
\tablenotetext{a}{Only given when known (Sample I).  Set to zero if unknown.}
\tablenotetext{b}{Assumes $M_{gas} = 1.4 M_{HI}$.}
\tablenotetext{c}{For $M/L = 1.0$ in the $R$-band.}
\tablecomments{The complete version of this table is in the electronic
edition of the Journal.  The printed edition contains only a sample.
These data are also available in electronic format from
http://www.atnf.csiro.au/$\sim$edeblok/data and
http://www.astro.umd.edu/$\sim$ssm/data.}
\end{deluxetable}

\begin{deluxetable}{lllrrrrrrrrrrr} 
\tablecolumns{8} 
\tablewidth{0pt}
\tablecaption{Comparison of fitting parameters\label{comptable}}
\tablehead{
\noalign{\vskip 2pt} 
\colhead{} & \colhead{} & \colhead{} & \multicolumn{5}{c}{NFW halo, minimum disk}\\
\cline{4-8} \\
\colhead{Galaxy} & \colhead{Obs} & \colhead{Curve} &
\colhead{$c$} & \colhead{$\Delta c$}& \colhead{$V_{200}$} &
\colhead{$\Delta V_{200}$} & \colhead{$\chi^2_{red}$}}
\startdata F563-V2
& SMT & SMT &16.2 & 3.4 & 84.5 & 10.4 & 2.516 \\
F563-V2 & SMT & dBMR &7.5 & 3.7 & 153.1 & 70.5 & 1.391 \\
F563-V2 & SMT & data &5.9 & 2.2 & 192.3 & 76.4 & 3.42 \\[5pt] 
F568-1 & SMT & SMT & 13.4 & 1.1 & 112.1 & 6.3 & 0.265 \\
F568-1 & SMT & dBMR & 6.4 & 2.5 & 194.6 & 71.9 & 0.804 \\
F568-1 & SMT & data & 8.3 & 1.1 & 154.5 & 18.5 & 3.49 \\[5pt] 
F568-3 & SMT & SMT & 5.1 & 2.9 & 160.3 & 88.0 & 2.147 \\
F568-3 & SMT & dBMR & \emph{1.17} & \nodata &\emph{591.0}& \nodata & 3.551\\
F568-3 & SMT & data & 1.71 & 0.5 & 400.4 & 94.8 & 13.9 \\[5pt]
F568-3 & Paper I & dBMR & 3.2& 3.7& 214.6& 233.9& 2.239 \\
F568-3 & Paper I & data\tablenotemark{a} & 4.6& 0.5& 168.4& 17.7 & 8.01\\[5pt] 
F568-V1 & SMT & SMT & 14.2 & 0.7 & 91.5 & 2.3 & 0.239 \\
F568-V1 & SMT & dBMR & 14.6 & 1.2 & 92.1 & 4.9 & 0.197 \\
F568-V1 & SMT & data & 15.8 & 1.1 & 85.7 & 3.8 & 12.7 \\[5pt] 
F574-1 & SMT & SMT & 9.4 & 0.7 & 91.2 & 4.3 & 0.421 \\
F574-1 & SMT & dBMR & 8.3 & 1.3 & 98.3 & 10.4 & 1.595 \\
F574-1 & SMT & data & 8.2 & 0.4 & 99.3 & 3.4 & 3.84 \\ \enddata
\tablecomments{Italics indicate estimates, not actual fits.  $V_{200}$
is in km~s$^{-1}$.  The column labeled ``Obs'' gives the source of the
raw data.  The column labeled ``Curve'' gives the source for the derived
rotation curve: ``SMT'' indicates smooth rotation curve from SMT;
``dBMR'' indicates smooth rotation curve from this paper; ``data''
indicates a fit to the raw data.} \tablenotetext{a}{Uncertain, depends on initial estimates of fit.} \end{deluxetable}

\begin{deluxetable}{lrrrrrrrrrrrrrr}
\tablewidth{0pt}
\tablecolumns{15}
\tablewidth{0pt}
\tablecaption{Fitting parameters NFW halo, Sample I\label{nfwtable}}
\tablehead{
\noalign{\vskip 2pt}
\colhead{}&\multicolumn{6}{c}{minimum disk}&\colhead{}&\multicolumn{6}{c}{minimum disk + gas}\\
\cline{2-7}\cline{9-14}\\
\colhead{Galaxy}&\colhead{$c$}&\colhead{$\Delta c$}&\colhead{$V_{200}$}&\colhead{$\Delta V$}&\colhead{$\chi^2_{red}$}&\colhead{$p$}&\colhead{}&\colhead{$c$}&\colhead{$\Delta c$}&\colhead{$V_{200}$}&\colhead{$\Delta V$}&\colhead{$\chi^2_{red}$}&\colhead{$p$}}
\startdata
F563-1 & 10.7  & 1.2 & 93.1 & 4.3 & 0.092   & 0.999 && 11.3  & 1.3  & 87.5  &  3.9  &  0.089  &  0.999&\\
F568-3 & 3.2   & 3.7 &214.6 &233.9& 2.239   & 0.017 && 4.2   & 3.3  &161.3  &118.9  &  2.386  &  0.011&\\
F571-8 & 7.8   & 1.1 &163.8 &20.2 & 1.501   & 0.123 && 7.8   & 1.1  &163.3  & 20.1  &  1.477  &  0.132&\\
F579-V1& 20.9  & 1.5 & 78.4 &  2.6& 0.211   & 0.998 && 22.1  & 1.6  & 75.1  &  2.5  &  0.217  &  0.998&\\
F583-1 &  5.1  & 1.0 &106.6 & 17.0& 0.740   & 0.746 && 6.2   & 1.1  & 86.6  & 12.4  &  0.827  &  0.648&\\
F583-4 &  5.7  & 1.4 & 89.5 & 19.0& 0.322   & 0.944 && 6.1   & 1.4  & 82.2  & 15.5  &  0.321  &  0.945&\\
U5750  &  2.6  & 1.5 &123.1 & 58.8& 1.243   & 0.262 && 2.9   & 1.6  &105.8  & 47.7  &  1.203  &  0.288&\\
U6614  & 10.3  & 2.0 &169.8 & 17.7& 4.626   & 0.000 && 11.0  & 2.2  &163.9  & 17.1  &  4.712  &  0.000&\\
\sidehead{\emph{SMT data, our analysis}}
F563-V2 & 7.5 & 3.7 & 153.1 & 70.5 & 1.391 & 0.195 && 8.3 & 3.8 & 133.1 & 54.1 & 1.484 & 0.157 & \\
F568-1  & 6.4 & 2.4 & 194.6 & 71.9 & 0.804 & 0.625 && 7.5 & 2.4 & 160.1 & 46.7 & 0.869 & 0.562 & \\
F568-3 &\emph{1.2}&\nodata&\emph{591.1}&\nodata&3.551&0.000&&\emph{1.2}&\nodata&\emph{552.6}&\nodata&3.573&0.000\\
F568-V1&   14.6& 1.2 &92.0  &  4.9& 0.197   &0.999  && 15.7  &1.5   &87.2   &  5.0  &  0.242  &  0.997&\\
F574-1&     8.3& 1.3 &98.3  & 10.4& 1.595   &0.085  &&  9.0  &1.5   &91.1   &  9.4  &  1.806  &  0.041&\\
\tableline
\noalign{\vskip 2pt}
&\multispan{6}{\hfill constant ${\MLstar}(R)=1.4$\hfill }&&\multispan{6}{\hfill maximum disk\hfill\ }\\[2pt]
\cline{2-7}\cline{9-15}\\
Galaxy&$c$&$\Delta c$&$V_{200}$&$\Delta V$&$\chi^2_{red}$&$p$&&$c$&$\Delta c$&$V_{200}$&$\Delta V$&$\chi^2_{red}$&$p$&$\Upsilon_*^R$\\[2pt]
\tableline
\noalign{\vskip 2pt}
F563-1 & 9.9  &1.2  & 88.8 & 4.6 & 0.089   & 0.999 && 4.0   & 1.2  &  110.0& 20.1  &  0.098  &  0.999  &6.9\\
F568-3 & 2.3  &4.9  &218.6 &410.9& 2.127   & 0.024 && 0.4   &19.5  &  595.0&$\infty$& 2.015  &  0.024  &2.2\\
F571-8 & 1.6  &5.7  &591.4 &\nodata&3.776 & 0.012 &&\emph{1.0}&\nodata&\emph{500.0}&\nodata&7.060&  0.000  &4.2\\
F579-V1&23.2  &1.9  & 67.9 &  2.4& 0.215   & 0.998 && 43.4  &14.6  & 31.6  & 3.9 & 0.671 & 0.781 &  7.9 \\
F583-1& 5.4&1.1&90.6& 14.7&0.767&0.716&&2.2&  1.5&145.5&78.7&0.680&0.804&6.5\\
F583-4&5.9&1.4&76.0& 14.5&0.271&0.965&&11.4&  6.1&12.9&  3.3&0.196&0.986&9.6\\ 
U5750&1.9& 1.7 &116.9& 80.4 &1.105&0.354&&1.9& 1.7 &116.9& 80.4 &1.105&0.354&1.4\\
U6614&1.7&1.8&303.0&206.5&4.005&0.001&&\emph{0.4}&   \nodata &\emph{145.6}&   \nodata &7.828&0.737&7.7\\
\sidehead{\emph{SMT data, our analysis}}
F563-V2&6.5&3.8&136.0&73.3&1.047&0.397&&\emph{1.0}&   \nodata &\emph{350.0}&   \nodata &0.449&0.878&4.1\\
F568-1&6.7&2.5&163.8& 56.1&0.803&0.626&&\emph{0.6}&   \nodata &\emph{669.2}&   \nodata &0.636&0.898&9.0\\
F568-3&\emph{1.0}& \nodata &\emph{519.4}&   \nodata &3.595&0.000&&\emph{1.0}&   \nodata &\emph{467.8}&   \nodata &3.624&0.000&1.8\\
F568-V1&14.6&1.4&88.8&  5.3&0.228&0.998&&3.3&  1.5&125.8& 45.7&0.222&0.998&14.0\\
F574-1&8.2&1.4&87.1& 9.5&1.391&0.162&&1.5&  1.6&69.7& 43.9&0.204&0.998&8.1\\ 
\enddata
\tablecomments{Italics indicate estimates, not actual fits.
$V_{200}$ is in km s$^{-1}$.}
\end{deluxetable}

\begin{deluxetable}{lrrrrrr}
\tablewidth{0pt}
\tablecolumns{7}
\tablewidth{0pt}
\tablecaption{Fitting parameters NFW halo, Sample II\label{nfwesotable}}
\tablehead{
\noalign{\vskip 2pt}
\colhead{}&\multicolumn{6}{c}{minimum disk}\\
\cline{2-7}\\
\colhead{Galaxy}&\colhead{$c$}&\colhead{$\Delta c$}&\colhead{$V_{200}$}&\colhead{$\Delta V$}&\colhead{$\chi^2_{red}$}&\colhead{$p$}}
\startdata
F730-V1 & 11.8 & 1.9 & 131.4  & 16.2 & 0.995 & 0.426\\
U4115   & \emph{5.0} &\nodata&\emph{133.4}&\nodata&0.777&0.591\\
U11454  & 10.4 & 2.0 & 152.6  & 23.3 & 3.334 & 0.000\\
U11557  & \emph{1.0} &\nodata&\emph{425.1}&\nodata&1.367&0.093\\
U11583  & \emph{5.0} &\nodata&\emph{93.3}&\nodata&0.676&0.641\\
U11616  & 12.7       &1.8    &   124.4    &  14.3 &1.254 &0.244\\
U11648  & 8.0        &0.7    &   146.2    &10.9   &0.964 &0.498\\
U11748  & 52.5       &4.5    &   125.2    &3.8    &3.325 &0.000\\
U11819  & 6.4        &1.9    &  252.9     & 77.6  &1.348 &0.177\\
E0140040&     16.8   &0.8    &     203.3  &  5.8  &0.152 &0.989\\
E0840411& \emph{1.0} &\nodata&\emph{181.0}&\nodata&1.608 &0.077\\
E1200211&     6.4    &2.1    &     27.5   &  6.9  &0.246 &0.996\\
E1870510&     3.8    &1.5    &     93.6  & 38.6  &0.059 &1.000\\
E2060140&     15.2   &1.3    &    92.7   &  4.4  &0.425 &0.962\\
E3020120&     6.6    &1.5    &   98.5     & 18.4  &0.333 &0.965\\
E3050090& \emph{1.0} &\nodata&  323.6 & \nodata & 0.208 & 0.999\\ 
E4250180&     3.1    &0.9    &  301.6     & 77.8  &0.014 &1.000\\
E488-049&     4.9    &1.9    &  209.6     & 90.1  &0.170 &0.999\\
\enddata
\tablecomments{Italics indicate estimates, not actual fits. $V_{200}$ is in km~s$^{-1}$.}
\end{deluxetable}

\begin{deluxetable}{lrrrrrrrrrrrrrr}
\tablewidth{0pt}
\tablecolumns{15}
\tablewidth{0pt}
\tablecaption{Fitting parameters ISO halo, Sample I\label{isotable}}
\tablehead{
\noalign{\vskip 2pt}
\colhead{}&\multicolumn{6}{c}{minimum disk}&\colhead{}&\multicolumn{6}{c}{minimum disk + gas}\\
\cline{2-7}\cline{9-14}\\
\colhead{Galaxy}&\colhead{$R_C$}&\colhead{$\Delta R$}&\colhead{$\rho_0$}&\colhead{$\Delta \rho$}&\colhead{$\chi^2_{red}$}&\colhead{$p$}&\colhead{}&\colhead{$Rc$}&\colhead{$\Delta R$}&\colhead{$\rho_0$}&\colhead{$\Delta \rho$}&\colhead{$\chi^2_{red}$}&\colhead{$p$}}
\startdata
F563-1 & 1.72 & 0.23 & 91.9 & 21.6 & 0.085 & 1.000 & & 1.55 & 0.22 & 102.0 & 25.2 & 0.078 & 1.000\\
F568-3  & 2.92 & 0.36 & 36.6 & 5.4 & 0.522 & 0.860 &  & 2.71 & 0.39 & 38.3 & 6.8 & 0.676 & 0.731\\
F571-8 & 2.12 & 0.19 & 106.9 & 14.0 & 1.525 & 0.114 & & 2.12 & 0.19 & 106.3 & 14.0 & 1.512 & 0.118\\
F579-V1 & 0.67 & 0.02 & 574.8 & 37.3 & 0.026 & 1.000 &  & 0.63 & 0.03 & 630.8 & 50.2 & 0.037 & 1.000\\
F583-1 & 2.44 & 0.06 & 33.0 & 1.1 & 0.037 & 1.000 & & 2.08 & 0.11 & 37.7 & 2.5 & 0.117 & 1.000\\
F583-4 & 1.10 & 0.13 & 85.5 & 15.8 & 0.329 & 0.941 & & 1.06 & 0.11 & 88.2 & 15.1 & 0.267 & 0.967\\
U5750 & 4.25 & 0.39 & 10.6 & 1.0 & 0.154 & 0.998 & & 3.96 & 0.49 & 10.4 & 1.3 & 0.214 & 0.993\\
U6614 & 1.86 & 0.49 & 218.4 & 102.5 & 1.942 & 0.021 & & 1.73 & 0.44 & 244.7 & 111.6 & 1.777 & 0.040\\
\sidehead{\emph{SMT data, our analysis}}
F563-V2 & 1.69 & 0.17 & 131.2 & 19.4 & 0.283 & 0.972 & & 1.58 & 0.20 & 135.3 & 24.7 & 0.393 & 0.925\\
F568-1 & 2.22 & 0.10 & 97.2 & 6.2 & 0.066 & 1.000 &  & 2.03 & 0.12 & 104.6 & 8.9 & 0.106 & 1.000\\
F568-3 & 3.93 & 0.75 & 30.2 & 5.6 & 1.256 & 0.245 &  & 3.75 & 0.78 & 30.8 & 6.3 & 1.383 & 0.173\\
F568-V1 & 1.45 & 0.11 & 153.0 & 18.2 & 0.110 & 1.000 &  & 1.33 & 0.14 & 170.0 & 27.8 & 0.172 & 1.000\\
F574-1 & 1.83 & 0.06 & 71.5 & 3.7 & 0.100 & 1.000 & & 1.70 & 0.09 & 76.9 & 6.3 & 0.232 & 0.997\\
\tableline
\noalign{\vskip 2pt}
&\multispan{6}{\hfill constant ${\MLstar}(R)=1.4$\hfill }&&\multispan{6}{\hfill maximum disk\hfill\ }\\[2pt]
\cline{2-7}\cline{9-15}\\
Galaxy&$R_C$&$\Delta R$&$\rho_0$&$\Delta \rho$&$\chi^2_{red}$&$p$&&$R_C$&$\Delta R$&$\rho_0$&$\Delta \rho$&$\chi^2_{red}$&$p$&$\Upsilon_\star^R$\\[2pt]
\tableline
\noalign{\vskip 2pt}
F563-1 & 1.72 & 0.26 & 79.0 & 21.3 & 0.083 & 1.000 & & 4.09 & 1.01 & 13.2 & 4.9 & 0.124 & 0.998 & 6.9\\
F568-3 & 3.07 & 0.63 & 25.7 & 6.3 & 0.793 & 0.623 &  & 3.36 & 0.88 & 19.7 & 5.9 & 0.89 & 0.536 & 2.2\\
F571-8 & 4.19 & 0.28 & 34.4 & 2.7 & 0.405 & 0.954 &  & 9.97 & 3.43 & 9.8 & 2.3 & 2.639 & 0.002 & 4.2\\
F579-V1 & 0.55 & 0.04 & 694.4 & 84.0 & 0.06 & 1.000 &  & 0.17 & 0.16 & 1970 & 3234 & 1.032 & 0.415 & 7.9\\
F583-1 & 2.26 & 0.11 & 31.5 & 2.0 & 0.103 & 1.000 & & 3.41 & 0.24 & 14.3 & 1.2 & 0.121 & 1.000 & 6.5\\
F583-4 & 1.02 & 0.12 & 80.6 & 15.2 & 0.249 & 0.972 &  & 0.23 & 0.15 & 103.4 & 114.1 & 0.188 & 0.988 & 9.6\\
U5750 & 4.67 & 0.74 & 7.1 & 1.1 & 0.262 & 0.984 &  & 4.67 & 0.74 & 7.1 & 1.1 & 0.262 & 0.984 & 1.4\\
U6614 & 12.18 & 2.87 & 6.3 & 1.9 & 1.938 & 0.022 &  & 112.0 & 506.7 & 0.4 & 0.4 & 4.762 & 0.000 & 7.4\\
\sidehead{\emph{SMT data, our analysis}}
F563-V2 & 1.70 & 0.24 & 96.6 & 19.8 & 0.298 & 0.967 & & 2.32 & 0.63 & 30.8 & 11.5 & 0.191 & 0.992 & 4.1\\
F568-1 & 2.11 & 0.15 & 90.7 & 9.0 & 0.120 & 1.000 & & 3.08 & 0.73 & 27.6 & 8.5 & 0.270 & 0.988 & 9.0\\
F568-3 & 4.35 & 1.31 & 21.5 & 6.0 & 1.709 & 0.065 & & 4.54 & 1.52 & 19.4 & 5.9 & 1.815 & 0.046 & 1.8\\
F568-V1 & 1.41 & 0.15 & 146.1 & 23.0 & 0.153 & 1.000 &  & 3.80 & 0.58 & 14.8 & 2.6 & 0.120 & 1.000 & 14.0\\
F574-1 & 1.74 & 0.09 & 63.1 & 5.4 & 0.182 & 0.999 &  & 3.30 & 0.83 & 4.6 & 1.6 & 0.108 & 1.000 & 8.1\\
\enddata
\tablecomments{$R_C$ is in kpc. $\rho_0$ is expressed in units of $10^{-3}\ M_{\odot}$ pc$^{-3}$.}
\end{deluxetable}

\begin{deluxetable}{lrrrrrr}
\tablewidth{0pt}
\tablecolumns{7}
\tablewidth{0pt}
\tablecaption{Fitting parameters ISO halo, Sample II\label{isoesotable}}
\tablehead{
\noalign{\vskip 2pt}
\colhead{}&\multicolumn{6}{c}{minimum disk}\\
\cline{2-7}\\
\colhead{Galaxy}&\colhead{$R_C$}&\colhead{$\Delta R$}&\colhead{$\rho_0$}&\colhead{$\Delta \rho$}&\colhead{$\chi^2_{red}$}&\colhead{$p$}}
\startdata
F730-V1		&1.46 & 0.06 & 215.8 & 14.2 &0.097 & 0.997 \\
U4115		&0.94 & 0.03 & 148.2 & 2.4  &0.004 & 1.000 \\
U11454		&1.95 & 0.10 & 146.9 & 11.8 &0.423 & 0.927\\
U11557		&5.48 & 0.55 & 15.2  & 0.9  &0.052 & 1.000\\
U11583		&0.63 & 0.08 & 117.8  & 16.5&0.103& 0.999\\
U11616		&1.45 & 0.05 & 208.9 & 11.6 &0.140& 1.000\\
U11648		&1.95 & 0.25 & 104.9 & 20.7 &3.792& 0.000\\ 
U11748		&0.36 & 0.15 & 8540 & 6661 &5.402& 0.000\\
 U11819 	&2.93 & 0.14 & 88.2  & 5.2  &0.303 &0.991  \\
 E0140040 	&2.55 & 0.18 & 249.5 & 27.0 &0.176 &0.982\\
 E0840411 	&6.41 & 0.56 & 5.2  & 0.3  &0.067 &0.999\\
 E1200211 	&0.57 & 0.08 & 45.5  & 9.2  &0.082 &1.000\\
 E1870510 	&0.97 & 0.05 & 53.5  & 3.2  &0.028 &1.000\\
 E2060140 	&1.17 & 0.05 & 231.1 & 16.9 &0.106 &1.000\\
 E3020120 	&1.90 & 0.09 & 53.6 & 3.4  &0.035 &1.000\\
 E3050090	&2.09 & 0.14 & 27.3 & 1.8  & 0.048& 1.000\\
 E4250180 	&4.41 & 0.75 & 30.0  & 6.6  &0.088 &0.997  \\
 E488-049 	&1.63 & 0.04 & 101.1  & 3.1  &0.016 &1.000\\
\enddata
\tablecomments{$R_C$ is in kpc. $\rho_0$ is expressed in units of $10^{-3}\ M_{\odot}$ pc$^{-3}$.}
\end{deluxetable}

\begin{deluxetable}{lrrrrr}
\tablewidth{0pt}
\tablecolumns{7}
\tablewidth{0pt}
\tablecaption{Comparison probabilities NFW vs.~ISO Sample I\label{veratable}}
\tablehead{
\noalign{\vskip 2pt}
\colhead{}&\multicolumn{5}{c}{Number of galaxies (out of 13)}\\
\noalign{\vskip 2pt}
\colhead{}&\multicolumn{2}{c}{ISO halo}&\colhead{}&\multicolumn{2}{c}{NFW halo}\\
\cline{2-3}\cline{5-6}
\noalign{\vskip 2pt}
\colhead{\MLstar}&\colhead{$p>0.95$}&\colhead{$p<0.05$}&\colhead{}&\colhead{$p>0.95$}&\colhead{$p<0.05$}}
\startdata
minimum	& 8 & 1 && 4 & 3 \\ 
min+gas	& 8 & 1 && 4 & 4 \\
constant& 10 & 1 && 4 & 4 \\
maximum	& 8 & 3 && 6 & 3 \\
\enddata
\tablecomments{$p$ is the probability that the model is compatible with the data.} 
\end{deluxetable}

\begin{deluxetable}{lllll}
\tablewidth{0pt}
\tablecolumns{5}
\tablewidth{0pt}
\tablecaption{Morphology and probability NFW halos\label{morphtable}}
\tablehead{
\noalign{\vskip 2pt}
\colhead{Galaxy}&\colhead{prob.\tablenotemark{a}}&\colhead{core\tablenotemark{b}}&\colhead{bar\tablenotemark{b}}&\colhead{morphology}}
\startdata
F563-1		& +	& --	& +  & mag.~irr. \\
F563-V2		& 0	& --	& +  & mag. bar	\\
F568-1		& 0	& +	& -- & spiral	\\
F568-3		& --	& --	& +  & spiral with mag. bar\\
F568-V1		& +	& +	& -- & spiral	\\
F571-8		& 0	& +	& ?  & edge-on	\\
F574-1		& --	& + 	& -- & disk	\\
F579-V1		& +	& +	& -- & core, flocc.~arms	\\
F583-1		& 0	& +	& -- & mag.~irr.	\\
F583-4		& +	& +?	& +? & fuzzy	\\
F730-V1		& 0	& +	& -- & spiral \\
U4115 		& 0     & +?    & -- & fuzzy \\
U5750		& 0	& --	& +  & mag.~bar	\\
U6614		& --	& +	& -- & faint, with bulge	\\
U11454		& --	& +	& -- & fuzzy spiral, small core	\\
U11557		& 0	& +	& -- & fuzzy spiral, small core	\\
U11583		& 0	& --	& +  & faint mag. bar	\\
U11616		& 0	& +	& --  & fuzzy, irr\\
U11648		& 0	& --	& +  & irr	\\
U11748		& --	& +	& +? & irr, bright core/bar?	\\
 U11819 	& 0	& +	& -- & fuzzy	\\
 E0140040 	& +	& +	& -- & spiral	\\
 E0840411 	& --	& --	& ?  & edge-on	 \\
 E1200211 	& +	& --	& +  & fuzzy mag.~bar	\\
 E1870510 	& +	& --	& +  & irr. spiral, flocc.\\
 E2060140 	& +	& +	& -- & spiral	\\
 E3020120 	& +	& +	& +? & spiral, hint of bar?	\\
 E3050090       & +     & +     & +  & barred spiral \\
 E4250180 	& +	& +	& -- & spiral	\\
 E488-049 	& +	& --	& +  & inclined mag.~bar	\\
\enddata
\tablenotetext{a}{``+'' indicates good fit $p\geq 0.95$; ``0'' indicates average fit $0.05 < p < 0.95$; ``--'' indicates bad fit $p< 0.05$.}
\tablenotetext{b}{``+'' indicates component clearly present; ``--'' indicates component not obviously present.}
\end{deluxetable}

\begin{deluxetable}{llll}
\tablewidth{0pt}
\tablecolumns{4}
\tablewidth{0pt}
\tablecaption{Morphology and probability NFW halos\label{morphstattable}}
\tablehead{\noalign{\vskip 2pt}
\colhead{quality }&\colhead{bar}&\colhead{core}&\colhead{both}}
\startdata
good ($p>0.95$)	&	4 & 7 & 3 \\
bad  ($p<0.05$)	&	2 & 4 & 0 \\
unclear ($0.05<p<0.95$)& 	4 & 7 & 1 \\
\enddata
\tablecomments{Value indicates number of minimum disk fits of that quality in presence of component mentioned.}
\end{deluxetable}

\end{document}